\documentclass[
reprint,
superscriptaddress,
 amsmath,amssymb,
 aps,
prfluids, onecolumn,
nofootinbib]{revtex4-2}

\usepackage[utf8]{inputenc}
\usepackage{amsmath,amssymb,bm}
\usepackage{graphicx,color}
\usepackage{lineno}
\usepackage{soul}
\usepackage{hyperref} 
\hypersetup{
    colorlinks=true,
    linkcolor=blue,
    citecolor=blue,      
    urlcolor=blue
    }

\newcommand{\boldsf}[1]{\boldsymbol{\mathsf{#1}}}

\newcommand{\Pe}{\mathrm{Pe}}
\newcommand{\nb}{\boldsymbol{n}}
\newcommand{\ub}{\boldsymbol{u}}
\newcommand{\vb}{\boldsymbol{v}}
\newcommand{\bv}[1]{\boldsymbol{#1}}
\newcommand{\bt}[1]{\boldsymbol{\mathsf{#1}}}

\newcommand{\Rb}{\boldsymbol{R}}
\newcommand{\Jb}{\boldsymbol{J}}
\newcommand{\Fb}{\boldsymbol{F}}
\newcommand{\qb}{\boldsymbol{d}}
\newcommand{\xb}{\boldsymbol{x}}
\newcommand{\rb}{\boldsymbol{r}}
\newcommand{\pb}{\boldsymbol{p}}

\newcommand{\mb}{\boldsymbol{m}}

\newcommand{\nA}{\bar{n}_\mathcal{A}}
\newcommand{\nB}{\bar{n}_\mathcal{B}}
\newcommand{\eq}[1]{#1^\mathrm{eq}}
\newcommand{\mA}{\bar{m}_\mathcal{A}}

\newcommand{\qbar}{\bar{\boldsymbol{d}}_\mathcal{A}}
\newcommand{\pard}[2]{\frac{\partial #1}{\partial #2}}

\newcommand{\bnabla}{\boldsymbol{\nabla}}

\begin{document}


\title{Pore-scale distribution and transport of active particles in a two-dimensional lattice}

\author{Akhil Varma}
\affiliation{Department of Mechanical and Aerospace Engineering, University of California San Diego, 9500 Gilman Drive, La Jolla, CA 92093, USA
}%
\affiliation{Max Planck Institute for the Physics of Complex Systems, 38 N\"othnitzer Str., Dresden, 01187 Germany}

\author{David Saintillan}%
\email{dstn@ucsd.edu}
\affiliation{Department of Mechanical and Aerospace Engineering, University of California San Diego, 9500 Gilman Drive, La Jolla, CA 92093, USA
}%


\begin{abstract}
Suspensions of motile microswimmers such as bacteria and other active colloids frequently encounter porous environments where obstacles and complex shear flows strongly influence their dynamics. Here, we study the distribution and transport of a dilute suspension of active particles in a square lattice of pillars, which serves as a model porous medium. The microswimmers are modeled as slender point particles, and Brownian Dynamics simulations are performed to determine how their number density and polarization fields change with systematic variations in the medium porosity, polydispersity, flow strength, and self-propulsion strength. We find that in the absence of flow, self-propulsion drives particle accumulation and radial polarization at the pillar surfaces. In the presence of a background flow, particles preferentially accumulate in the wake of pillars and exhibit upstream polarization near their surface, consistent with experimental observations. At moderate flow strengths, topological defects nucleate in the polarization field. These defects are of purely kinematic origin and mark the transition from global upstream swimming at low flow strengths to the coexistence of upstream and downstream swimming regions in the lattice at high flow strengths. The structured lattice studied here provides a controlled framework for isolating the physical mechanisms governing active transport in complex geometries, with direct relevance to transport in structured microfluidic settings.
\end{abstract}

\maketitle

\vspace*{-0.6cm}
\section{Introduction}

Microswimmers are colloidal particles that self-propel in an aqueous medium and fundamentally represent an out-of-equilibrium system. They can be of biological origin (microbes such as bacteria, spermatozoa, etc..) or synthetic (such as catalytic or magnetic colloids). Some of these microorganisms play a central role in natural processes such as bio-remediation and in industries such as pharmaceuticals, bioreactors, and filters. Synthetic microswimmers have also shown an immense potential in recent years for medical diagnostics and targeted drug delivery. Microswimmers predominantly inhabit patterned and complex environments such as soil, coral reefs, and even in the body vasculature, crowded tissues and the intestine. The complexity of these environments arises from the presence of solid inclusions, non-uniform background flows, external force fields, toxins and other chemicals, and interactions with other microorganisms. The passive as well as active response of microbial communities to these environmental cues have been a major interest of study in the past decade for physicists, ecologists and biologists alike \cite{Bechinger16,Conrad18,Wheeler19}.

Colonization of surfaces by microswimmers is generally the first step in many biophysical processes like the formation of biofilms. A purely kinematic explanation credits rigid walls to block the persistent motion of the motile particles, causing their retention at the surface over a characteristic time associated with their rotational diffusivities \cite{Li11,Thery21,Mok19}. The suspension reaches a dynamical steady state with a higher number density of particles at the boundary \cite{YanBrady15,Ezhilan15,Ezhilan15b}. 
Such spontaneous accumulation of constituent particles at boundaries is a hallmark of active (non-equilibrium) suspensions. It is hence not surprising that there exists considerable literature on the topic, with one of the earliest biological observations made in the 1960s involving spermatozoa \cite{Rothschild63}.
Being in an aqueous medium, hydrodynamics also plays an important role in the collective dynamics and transport in active suspensions, and has been reviewed in detail by many authors \cite{Pedley92,Lauga16,SaintillanShelley13,Saintillan18}. Both the self-induced flow \cite{Berke08,Sipos15,Takagi14,Spagnolie15} as well as the background flow \cite{Secchi20,Mino18,Lee21,Ezhilan15} influence surface accumulation of microswimmers. In the present work, we shall however ignore the former (assuming the suspension to be so dilute that these interactions are negligible) and only analyze the effect of an externally-imposed background flow. Experiments have shown that a pressure-driven flow through a channel does not influence a uniform distribution of passive Brownian particles as long as their volume fraction is small; at higher volume fractions the particles migrate towards the centerline of the channel where the shear rates are lower \cite{Leighton87,Frank03}. In contrast, active particles migrate towards the walls and get entrapped in  high-shear regions \cite{Rusconi14,Ezhilan15}. Furthermore, in the presence of moderate flows, they have been observed to swim upstream near the walls (positive rheotaxis) \cite{Hill07,Mathijssen19}. Elongated bodies such as prolate spheroidal particles (which many bacteria are a good approximation of) undergo tumbling trajectories in which they spend more time aligned with the flow \cite{Jeffery1922,Bretherton62,Junot19}. It is this passive aligning effect from the flow, combined with active self-propulsion, that enables surface rheotaxis in many rod-shaped bacteria such as \textit{E.coli} and \textit{Salmonella} \cite{Kaya12}. Similar rheotactic behavior has also been observed in catalytic colloids \cite{Palacci15,Sharan22}. 

The above examples show that even the presence of a flat boundary or a simple shear flow yields non-trivial dynamics in motile colloidal suspensions. Bacteria and other microswimmers usually inhabit much more complex porous environments, and we are yet to fully understand the role of intricate networks of shear flow profiles and heterogeneous boundaries on transport through such media \cite{Creppy19,Amchin22,Das25}. To reduce the complexity of the problem, it is instructive to first look at the special case of an ordered lattice.  Experiments and particle simulations of motile cells in a regular lattice identify a long-time sub-diffusive behavior of the particles, attributed to the physical constraints in their persistent motion \cite{Zeitz17,Bruny19,Dehkharghani23}. 
Thus, the long-time dispersion of active colloids through the medium differs fundamentally from that of passive colloids \cite{AlonsoMatilla19,Peng20,Saintillan23,Mattingly25}.

Active colloidal suspensions are modeled either in the framework of individual particles (microscopic scale) or at the level of the suspension (mesoscopic scale). In microscopic models, each particle is defined by an instantaneous spatial location and an intrinsic direction along which it self-propels with some speed (termed as `activity'), which evolve over time based on their interactions with the surrounding environment. The Active Brownian Particle (ABP) model is one such model which additionally takes into account the background noise, and is frequently used to study active matter systems. Numerical simulations of the model often consider the particles to be of point size to determine their distribution statistics in a dilute suspension. If the particle size is comparable to pore size, phase separation and clogging of the medium can occur \cite{Reichhardt21,Das20}; the distribution statistics in this case is dictated by the shape of the particle as well \cite{Chen21}. 
An alternate approach, which models the suspension at a mesoscopic scale, is a continuum kinetic theory in which the instantaneous configuration of a suspension is described by a probability distribution function \cite{SaintillanShelley08a,Subramanian09}. This framework is used for computing the dynamics of dilute suspensions to determine the suspension rheology \cite{Saintillan18}. It allows one to explicitly determine some of the fundamental statistical properties of the suspension described by the moments of the probability distribution function --- such as the number density, polarization, nematic order and higher-order moments \cite{SaintillanShelley08a}. In channel flow, these models convincingly predict the shear-induced migration of active particles at the walls seen in experiments \cite{Ezhilan15,Bearon15,Vennamneni20}.
   
An interesting feature of active matter systems is their traveling wave-like nature even though they operate in the overdamped limit. This is due to their short-term ballistic nature resulting from their activity, which eventually degenerates as rotational diffusion decorrelates the particle orientations over time \cite{Dulaney20,Kurzthaler18}. Inter-particle interactions and confinement can amplify these density waves \cite{Simha02,Baskaran08,Lefauve14}. Bacteria and synthetic colloidal suspensions have been shown to exhibit these waves in experiments \cite{Takatori16,Geyer18,Kurzthaler18}.   
However, the literature on active matter waves is sparse and studies are ongoing.

As one can see from the many contemporary works highlighted above, there is currently an immense interest in the scientific community --- from statistical physics to medicine and technology --- to characterize the transport of active colloids through complex environments (see \cite{Bechinger16} for a review). Recent efforts have shed light on the steady-state distribution as well as the long-time dispersive transport of active particles in a porous medium \cite{YanBrady15,AlonsoMatilla19,Kjeldbjerg22,Saintillan23}, but the temporal evolution of the statistics of the suspension has largely been ignored. It is known that rigid boundaries act as entropic barriers to active particles and so, analyzing their short-time dynamics helps one understand the kinematics of accumulation. To our understanding, a systematic study of the role of the governing parameters viz. the particle activity, system confinement, as well as pillar geometry on the suspension statistics is missing. Furthermore, the particles are often found in saturated environments with frequent interstitial flows; how does the presence of flow modify the surface accumulation? We seek to answer these questions in the present work, where we investigate, through extensive numerical simulations, the dynamics of a suspension of active colloids in a periodic lattice.

The paper is organized as follows. In Sec.~\ref{sec:kinmodel}, we summarize the ABP model for the dynamics of a single active colloid. The corresponding continuum kinetic framework is also detailed in this section.  In Sec.~\ref{sec:single_unbounded}, we solve the continuum kinetic equations along with appropriate closure and boundary conditions numerically to determine the time evolution of a homogeneous suspension in the presence of an isolated pillar. Particular attention is paid to understand the time evolution of the number density and the polarization of the particles on the surface of the pillar. While this framework could also be used to determine the suspension dynamics in a doubly-periodic lattice of pillars (for example, see \cite{AlonsoMatilla19}), it is computationally difficult especially when dealing with multiple pillars, strong confinement and heterogeneity. A more pragmatic approach, with a foresight on analyzing more complicated geometries and heterogeneous environments in future works, is to use Brownian Dynamics (BD) simulations to determine the dynamics of a large number of ABPs; the statistical properties of the suspension are then evaluated from the local distribution of particles in the domain. We discuss the results of BD simulations performed in a square lattice in Sec.~\ref{sec:periodic}. In particular, the effect of varying the confinement of the lattice and the activity of the particles on the dynamics are discussed in Sec.~\ref{sec:porosity} and Sec.~\ref{sec:activity}, respectively. 
The simulations show surprising transient oscillations in the statistical properties of the suspension on the surface of the pillars at short times, which are characterized here. Porous media are usually polydisperse, and so, the effect of multiple pillar sizes in the lattice is also briefly discussed in Sec.~\ref{sec:polydispersity}. Finally, in Sec.~\ref{sec:flow}, we analyze in detail how a background flow through the lattice modifies the transport and distribution of the particles in the medium. Some concluding remarks and future directions are also provided in Sec.~\ref{sec:conclusions}.

\section{Modeling dilute active suspensions}
\label{sec:kinmodel}

Consider a dilute suspension of active particles in a 2D fluid domain. The particles are assumed to be point-sized and non-interacting with each other. The particles move in a 2D space described by a vector $\rb$. At any instant $t$, let the position of a particle in this space be given by $\Rb(t)$. All particles are assumed to have the same {constant} swim speed $v$, but each of them has an independent instantaneous orientation $\pb(t)$; here, $\pb=(\cos \theta,\sin \theta)$ is a unitary director with $\theta(t)$ being the {instantaneous} orientation angle. Being in microscopic scales, the particles are influenced by the thermal noise in the surrounding fluid and hence, they additionally undergo Brownian motion. For this reason, each particle effectively has a translational diffusivity $D_T$ and a rotational diffusivity $D_R$. 

Two complementary approaches are usually taken to model the dynamics of active suspensions viz. at a particle-scale, or at the (mesoscopic) scale of the suspension, both of which are discussed below. 

\subsection{Active Brownian Particle model}
\label{sec:ABPmodel}

The Active Brownian Particle (ABP) model is a particle-scale model that describes the instantaneous dynamics of an individual active particle in a viscous fluid. It is expressed as the overdamped Langevin equations,
\begin{align}
    \dot{\Rb}(t) = v\, \pb(t) + \sqrt{2 D_T}\, \boldsymbol{\xi}_T(t), \quad \mathrm{and} \quad
     \dot{\pb}(t) = \sqrt{2 D_R}\, \boldsymbol{\xi}_R(t) \times \pb(t).
\end{align}
Here, $\boldsymbol{\xi}_{T}$ and $\boldsymbol{\xi}_{R}$ denote independent Gaussian white noise vectors, each with zero mean and unit variance, i.e. $\langle \boldsymbol{\xi}_T(t) \rangle = \langle \boldsymbol{\xi}_R(t) \rangle = \boldsymbol{0}$ and $\langle \boldsymbol{\xi}_T(t)\, \boldsymbol{\xi}_T(t')^{\top} \rangle = \langle \boldsymbol{\xi}_R(t)\, \boldsymbol{\xi}_R(t')^{\top} \rangle = \boldsf{I}\, \delta(t-t')$, where $\boldsf{I}$ is the identity matrix. Assuming infinite dilution, we ignore the effect of any interactions between the particles on the dynamics.

We assume the particles move through a two-dimensional square lattice of solid circular inclusions of radius $a$ and lattice spacing $\ell$. The problem involves two additional length scales: the persistence length of the active particles $\ell_p=v/D_R$, and a diffusive length $\smash{\ell_d=\sqrt{D_T/D_R}}$. To analyze particle dynamics at the scale of the inclusions, we set the pillar radius $a$ as the characteristic length scale for our problem. We additionally choose the characteristic time as the reorientation time $D_R^{-1}$. Upon rescaling using the characteristic scales, the governing equations become
\begin{align}
    \dot{\Rb}^*(t^*) = \Pe_s\, \pb(t^*) + \sqrt{2 \kappa^2}\, \boldsymbol{\xi}_T(t^*),\quad \mathrm{and} \quad  \dot{\pb}(t^*) = \sqrt{2}\, \boldsymbol{\xi}_R(t^*) \times \pb(t^*),
    \label{eq:BDdimensionless}
\end{align}
where the superscript $^*$ denotes dimensionless quantities; we shall, however, drop this superscript in the rest of the manuscript for brevity. Two dimensionless parameters govern the time evolution and statistics of $\boldsymbol{R}$ and $\boldsymbol{p}$: (i) the ratio of the persistence length to the pillar size, known as the swim P\'eclet number, given by $\Pe_s=\ell_p/a = v/(a D_R)$; and (ii) the ratio of the diffusive length scale to pillar size, given by $\kappa=\ell_d/a = \sqrt{D_T/ (a^2 D_R)}$. The choice of these parameters dictates the dynamics of the particles. For swimming microorganisms, typical values of translational and rotational diffusivities are $D_T \sim 10^{-1}-10^{-2}\,\mu \mathrm{m}^2/\mathrm{s}$ and $D_R \sim 10^{-1}-10^{-3}\,\mathrm{rad}/\mathrm{s}^2$ with swimming speeds of $v \sim 10-100\, \mu \mathrm{m}/\mathrm{s}$. Taking these numbers into consideration, and noting that the inclusions are typically of size $a\sim 10-100\, \mu \mathrm{m}$, we have $\kappa^2 \ll 1$ in a porous medium. Following \cite{AlonsoMatilla19}, we fix $\kappa^2=0.1$ in all our simulations.
This modeling framework will be used later in Sec.~\ref{sec:periodic} for Brownian Dynamics simulations.

\subsection{Continuum kinetic model}
\label{sec:continuumModel}

When describing the global statistical configuration of a large number of  particles in a system, it is often convenient and more fruitful to use a continuum approach. Instead of solving for the dynamics of individual particles using Eq.~\eqref{eq:BDdimensionless}, one can alternatively describe the system in a macroscopic sense by introducing the local probability distribution of particles in the spatial and orientational domain \cite{DoiEdwards,SaintillanShelley08a}. Let the probability density of finding a particle at position $\rb$ having an orientation $\pb$ at some time $t$ be denoted by $\psi(\rb,\pb,t)$. The probability density function $\psi(\rb,\pb,t)$ fully characterizes the distribution of particles in the system at any instant. Net conservation of particle number dictates the normalization condition\begin{equation}
    \frac{1}{A_\mathcal{F}}\int_S \int_\Omega \psi(\rb,\pb,t)\;\mathrm{d}\pb\;\mathrm{d}\rb = \bar{n},
\end{equation}
where $\bar{n}$ is the mean number density of swimmers in the area $A_\mathcal{F}$ of the 2D fluid domain; $\Omega$ is the orientational domain, which is the unit circle in 2D. Neglecting inter-particle interactions, the conservation of $\psi(\rb,\pb,t)$ is governed by the dimensionless Fokker-Planck equation \cite{DoiEdwards}
\begin{align}
    \pard{\psi}{t} + \bnabla \cdot \boldsymbol{J} + \bnabla_p \cdot  \boldsymbol{J}_p = 0,
\label{eq:FP}        
\end{align}
where $\boldsymbol{J}= \Pe_s \pb\; \psi - \kappa^2 \bnabla \psi$ and $\boldsymbol{J}_p = - \bnabla_p \psi$ are the translational and orientational fluxes, respectively, and capture the same dynamics as the Langevin equations \eqref{eq:BDdimensionless}. Here, $\bnabla=\partial/\partial \rb$ is the spatial gradient operator and $\bnabla_p = (\bt{I}-\bv{pp}) \cdot \partial/\partial \pb$ is the gradient on the unit circle $\Omega$. A similar evolution equation of the probability density function for run-and-tumble particles was derived in \cite{Subramanian09}.

In solving for $\psi(\rb,\pb,t)$ in \eqref{eq:FP}, it is a standard procedure to decompose the function into its various orientational moments \cite{SaintillanShelley08a}, which have simple physical interpretations. Introducing $ \langle h(\pb)\rangle = \int_\Omega h(\pb)\; \psi(\rb,\pb,t)\; \mathrm{d}\pb$ as the orientational average, the zeroth, first and second moments, defined as
\begin{equation}
    n(\rb,t) = \langle 1 \rangle,\quad \mb(\rb,t) = \langle \pb \rangle,\quad \bt{Q}(\rb,t) = \langle \pb\pb - \bt{I}/2 \rangle,
\end{equation}
describe the local number density, polarization vector and nematic order tensor, respectively. 
Evolution equations for these fields can be obtained by taking orientational moments of the Fokker-Planck equation (\ref{eq:FP}), and form an infinite set of coupled tensorial equations that cannot be truncated at any order without the use of a closure model \cite{SaintillanShelley08a}. The hierarchy can be closed by noting that far from boundaries the higher-order moments relax more rapidly under rotational diffusion, and so, one needs to consider only the first few moments. The simplest closure that captures salient dynamics near walls consists in  setting the nematic order to zero, $\boldsf{Q}=\mathbf{0}$ \cite{YanBrady15}. Particle simulations indeed indicate that, for suspensions confined in a channel, nematic ordering is negligible over most of the domain \cite{Ezhilan15}. Other closure conditions are possible depending on the problem (see \cite{SaintillanShelley13} and references therein). Using the condition that $\boldsf{Q}=\mathbf{0}$, one obtains a closed system of coupled equations for the number density and polarization fields \cite{SaintillanShelley08a,SaintillanShelley08b},
\begin{align}
    \pard{n}{t} + \bnabla \cdot \bv{j}_n = 0, \quad \quad
    \pard{\mb}{t} + \bnabla \cdot \boldsf{j}_m + \mb = \mathbf{0},
    \label{eq:pdenm}
\end{align}
where the fluxes in 2D are given by
\begin{align}
    \bv{j}_n =  \Pe_s \mb - \kappa^2 \; \bnabla n, \quad \textrm{and} \quad \boldsf{j}_m = \frac{\Pe_s n}{2} \boldsf{I} - \kappa^2 \;\bnabla \mb.
    \label{eq:fluxnm}
\end{align}
In the following sections, we shall use the modeling approaches of Sec.~\ref{sec:ABPmodel} and \ref{sec:continuumModel} to describe the distribution of particles around an isolated pillar and in a square lattice of pillars.

\section{A pillar in an unbounded domain}
\label{sec:single_unbounded}

In this section we examine how the number density and polarization fields evolve over time in the presence of a solid circular boundary located at the origin in an otherwise unbounded suspension. We assume an initially homogeneous and isotropic suspension of particles of number density $n_\infty$, i.e. 
\begin{equation}
    n(\rb,0) =n_{\infty},\qquad \mb(\rb,0) = \boldsymbol{0}.
    \label{eq:initialnm}
\end{equation}
On the surface of the pillar ($\rb=\rb_s$), we have the no-flux boundary conditions,
\begin{align}
    \nb \cdot \bv{j}_{n}(\rb_{\mathrm{s}},t) = 0, \qquad \nb \cdot \boldsf{j}_m(\rb_{\mathrm{s}},t) =\mathbf{0},
    \label{eq:bc1nm}
\end{align}
where the fluxes $\bv{j}_{n}$ and $\boldsf{j}_m$ are defined in \eqref{eq:fluxnm}. Moreover, far from the pillar, the particles are unaffected by its presence and therefore the suspension maintains its homogeneous and isotropic state, i.e.
\begin{equation}
    n (r\to \infty,t) = n_\infty, \qquad \mb(r\to \infty,t) = \boldsymbol{0}.
    \label{eq:bc2nm}
\end{equation}
where $r=|\rb|$.

\begin{figure}
\centering
\includegraphics[width=\textwidth]{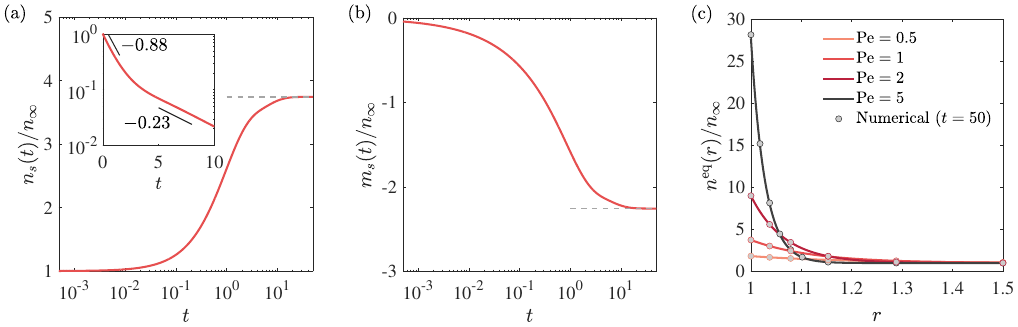}
\caption{Numerical solution for the time evolution of (a) the number density, $n/n_\infty$, and (b) the radial polarization, $m/n_\infty$, of the particles on the surface of a circular pillar of radius $r=1$ in an unbounded domain. The steady-state solution (Eqs. \eqref{eq:YanBradyA} and \eqref{eq:YanBradyB}) is indicated by a gray dashed line. Inset: normalized surface number density $\left(n_s(t)-\eq{n}_s\right)/(n_\infty-\eq{n}_s)$ highlighting exponential growth, {where the slope of the curve identifies the exponents, $k_1=0.88$ at short times and $k_2=0.23$ at long times, with a cross-over at $t\sim 1$}. Parameters used in (a) and (b) are $\Pe_s=1,\;\kappa^2=0.1$ (c) Comparison of the analytical steady-state number density across the domain (Eq. \eqref{eq:YanBradyA}) with the numerical solution at a long time, $t=50$.}
\label{fig:Brady}
\end{figure}

The coupled partial differential equations for $n$ and $\mb$ in \eqref{eq:pdenm}, along with the initial and boundary conditions in \eqref{eq:initialnm}-\eqref{eq:bc2nm}, are solved numerically using the method of lines (see Appendix \ref{app:wave} for more details). In the case of a circular pillar of unit radius, the symmetry of the problem allows one to seek a solution of the form $n(r,t) =n_{\infty} f(r,t)$ and $\mb(r,t) = n_\infty \rb g(r,t)$. Noting that the polarization field is purely radial, it is simply denoted by the scalar $m^\perp (r,t)=\mb(r,t) \cdot \hat{\rb}$, with $\hat{\rb}=\rb/r$ being the unit outward radial vector.  We observe in Fig. \ref{fig:Brady}($a,b$) that both the number density and the radial polarization of the particles on the surface of the pillar ($r=1$) increase over time, eventually reaching a steady state at long times. Note that the radial polarization is negative, implying a preferential alignment of particles towards the surface.
Surface accumulation is a characteristic feature of active suspensions, as self-propulsion drives them towards the wall faster than diffusion can redistribute them \cite{Bechinger16,Ezhilan15,YanBrady15}. 

Following \cite{YanBrady15}, solving for the steady-state ($\partial{\bullet}/\partial{t} = 0$) in {the two-moment equations} \eqref{eq:pdenm} gives analytical expressions for the long-time equilibrium number density and polarization fields as 
\begin{subequations}\label{eq:YanBrady} 
    \begin{align}
        \frac{\eq{n}(r)}{n_\infty} & = 1+\frac{2\, (\Pe_s/\kappa)^2\, K_0(r/\delta)}{K_0(1/\delta)\,(2-(\Pe_s/\kappa)^2) + K_2(1/\delta)\,(2+(\Pe_s/\kappa)^2)}, \label{eq:YanBradyA} \\
        {\frac{m^{\perp,\mathrm{eq}}(r)}{n_\infty}} & = \frac{-2\,(\Pe_s/\delta)\, K_1(r/\delta)}{K_0(1/\delta)\,(2-(\Pe_s/\kappa)^2) + K_2(1/\delta)\,(2+(\Pe_s/\kappa)^2)},
        \label{eq:YanBradyB} 
    \end{align}
\end{subequations}
where $r\geq 1$, and $K_i$ denotes the $i$-th order modified Bessel function of the second kind. A length scale $\delta$ naturally arises in the solution and characterizes the thickness of the accumulation layer on the surface of the pillar. The expression for $\delta$ in \eqref{eq:YanBrady} is given by
\begin{equation}
\frac{1}{\delta} = \frac{1}{\kappa}\sqrt{1+\frac{\Pe_s^2}{2\kappa^2}}.
\label{eq:delta}
\end{equation}
The distribution in Eq.~\eqref{eq:YanBradyA} is shown in Fig.~\ref{fig:Brady}(c). By increasing the activity $\Pe_s$, the number density on the surface increases, accompanied by sharper gradients. For large $r/\delta$, Eqs.~\eqref{eq:YanBradyA} and \eqref{eq:YanBradyB} both approximate to an exponential decay from the surface. We validated our numerical solutions by comparing the long-time distributions of $n(r,t)$ and $m(r,t)$ against these steady-state solutions for various $\Pe_s$ [see Fig. \ref{fig:Brady}(c)]. 

We now seek to characterize the growth of $n$ and $m$ on the surface at $r=1$, denoted by $n_s$ and $m_s$, from their initial values to their long-time equilibrium values in \eqref{eq:YanBrady}. The linearity of the coupled PDEs \eqref{eq:pdenm} suggests an exponential evolution towards equilibrium. As shown in Fig.~\ref{fig:Brady}(a, inset), the numerical solution indeed follows the ansatz 
\begin{equation}
    {\frac{n_s(t)-n^\mathrm{eq}_s}{n_\infty-n^\mathrm{eq}_s} = n_1 e^{-k_1 t} + n_2 e^{-k_2 t}} ,
    \label{eq:exponentialGrowth}
\end{equation}
with two distinct growth rates $(k_1, \; k_2)>0$ governing the dynamics at short and long times; a similar expression with the same exponents can be written for $m_s(t)$.

To facilitate comparison with discrete particle simulations in the following discussion, a more convenient measure of the surface accumulation is the number density averaged within the boundary layer $\mathcal{A}$ {of thickness $\delta$}, which is
\begin{equation}
\frac{\nA(t)}{n_\infty} = \frac{\int_\mathcal{A} n(r,t)/n_\infty \; \mathrm{d}\mathcal{S}}{A} = \frac{2 \int_1^{1+\delta} n(r,t)/n_\infty\; r \; \mathrm{d}r}{\delta(\delta+2)},
\label{eq:BLmean}
\end{equation}
where $A=\pi \delta(\delta+2)$ is the area within the boundary layer. Here, and in the remainder of the manuscript, we shall use $\bar{\bullet}_\Omega\;$ to denote the surface averaging operator in some specified closed domain $\Omega$. The analytical expression for \eqref{eq:BLmean} is provided in Eq.~\eqref{eq:blavg} of Appendix \ref{app:approx}. 
As an example, when $\Pe_s=1$ and $\kappa^2=0.1$, the steady state number density and radial polarization on the surface of the pillar are $\eq{n}_s/n_\infty=3.74$ and $m^\mathrm{eq}_s/n_\infty = -2.26$ respectively. These values are shown by the dashed lines in Fig. \ref{fig:Brady}(a,b). When averaged across the boundary layer (thickness, $\delta = 0.129$), the quantities are $\bar{n}^\mathrm{eq}_\mathcal{A}/n_\infty=2.67$ and $\bar{m}^\mathrm{eq}_\mathcal{A}/n_\infty = -1.37$.





\section{Periodic square lattice}
\label{sec:periodic}

\begin{figure}
    \centering
    \includegraphics[width=0.55\textwidth]{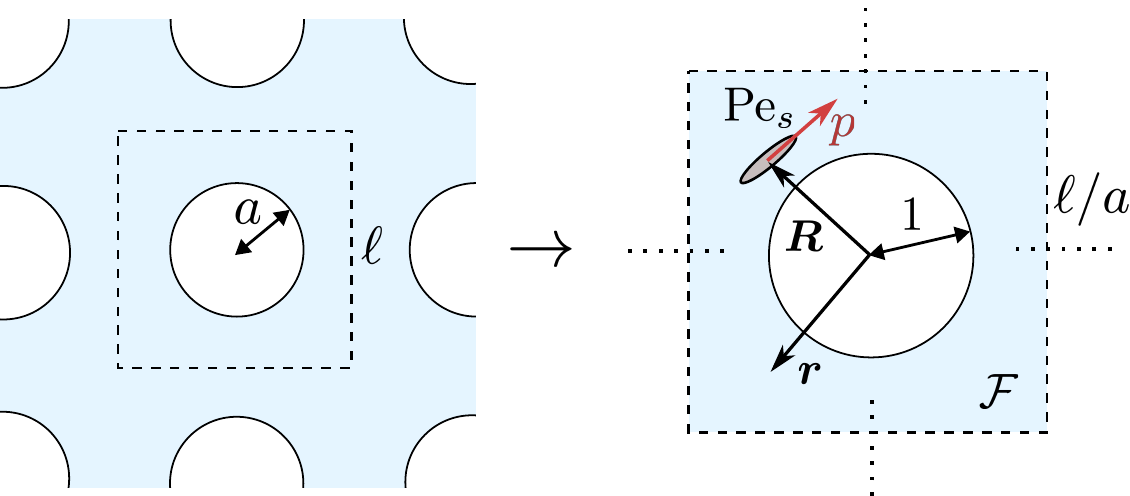}
    \caption{Schematic of (left) a square lattice formed by repeating a  cell of side $\ell$ containing a pillar of radius $a$ and (right) the dimensionless doubly-periodic cell used for numerical simulations. An Active Brownian Particle (ABP) in the interstitial region $\mathcal{F}$ is shown at some instantaneous position $\Rb(t)$, having an orientation $\boldsymbol{p}(t)$ and a fixed speed $\Pe_s$.}
    \label{fig:schematic}
\end{figure}

We now explore how the probability distribution of particles, in particular $n$ and $\mb$, change in a periodic lattice as opposed to a free domain. Specifically, we consider an infinitely large square lattice of pillars of radius $a$ with lattice spacing $\ell$, as illustrated in Fig. \ref{fig:schematic}(left). 
The porosity of the medium is defined as the fractional area of the fluid domain,
\begin{equation}
    {\epsilon = 1- \frac{\pi a^2}{\ell^2}}.
\end{equation}
The maximum radius of a pillar without any overlap is $a=\ell/2$, which gives a minimum porosity in this case of $\epsilon_\mathrm{min}=1-\pi/4 \approx 0.215$. Thus, the porosity of the medium $\epsilon \in [0.215,1)$, where $\epsilon=1$ denotes the limit of an unbounded domain. 

As before, the problem is made dimensionless by rescaling lengths with the pillar radius, $a$. Furthermore, symmetry and periodicity of the lattice reduce the problem to a doubly-periodic cell with a pillar of unit radius at its center (see right panel of Fig.~\ref{fig:schematic}). While numerical approaches such as finite volume or difference methods could be employed here to determine the distribution of the various moments \cite{AlonsoMatilla19}, computations become increasingly difficult with more complex pillar geometries or multiple pillars and under strong confinement. Instead, we perform Brownian Dynamics (BD) simulations based on the Langevin equations of Sec.~\ref{sec:ABPmodel} to determine the statistical distribution of the active particles in the medium. The interstitial domain is initially populated with a uniform distribution of $N$ point-sized ABPs, each with a random orientation, and the system is evolved in time using Eq.~\eqref{eq:BDdimensionless} (see Appendix \ref{app:B} for details). The particles do not interact with each other, but interact sterically with the pillars. The mean number density {and polarization} of the particles in the cell are
\begin{equation}
    \bar{n}=\frac{\int_\mathcal{F} n(\rb,t)\;\mathrm{d}S}{\int_\mathcal{F}\mathrm{d}S} = \frac{N}{A_\mathcal{F}}, \qquad  \bar{\mb}(t)=\frac{\int_\mathcal{F} \mb(\rb,t)\;\mathrm{d}S}{\int_\mathcal{F}\mathrm{d}S}, 
    \label{eq:meanDensityPolarization}
\end{equation}
where
\begin{equation}
A_\mathcal{F}={\epsilon\ell^2/a^2} = \pi\epsilon/(1-\epsilon)
\end{equation}
is the dimensionless area of the interstitial region {$\mathcal{F}$}. Here, $N$ is chosen sufficiently large to ensure statistical convergence. Particle conservation in the doubly-periodic cell implies $\bar{n}$ is fixed at all times; this is equivalent to assuming homogeneity across cells.

\begin{figure}[t]
\centering
\includegraphics[width=\textwidth]{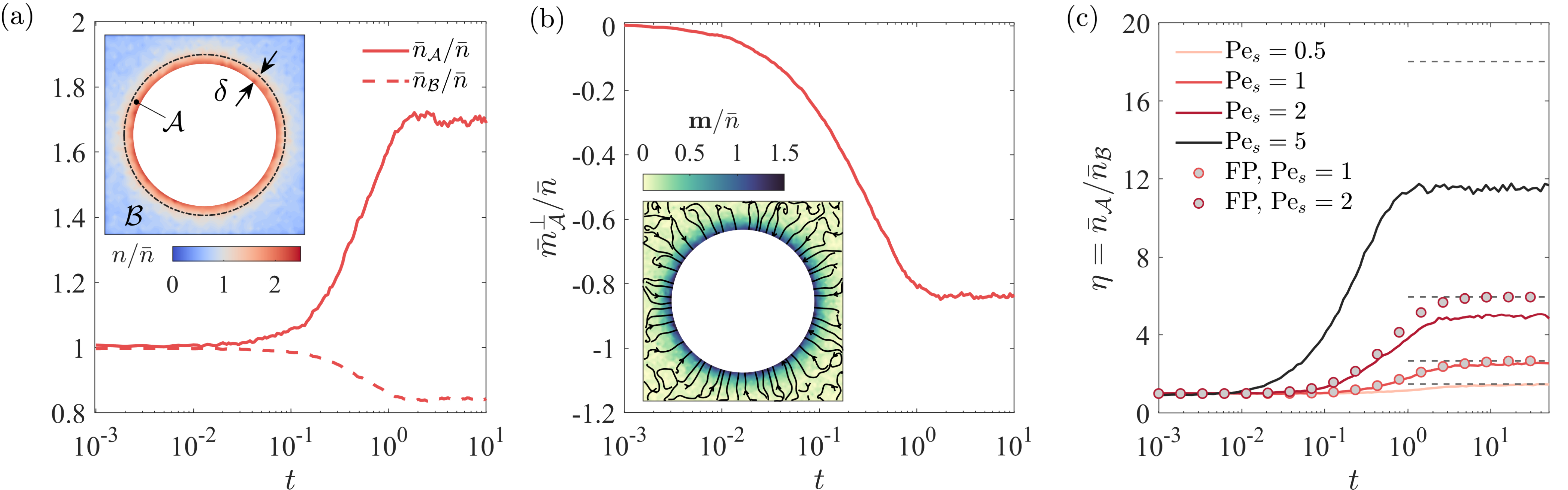}
\caption{Time evolution of (a) the mean number density in the boundary layer, $\nA(t)/\bar{n}$, and in the bulk, $\nB(t)/\bar{n}$, and (b) the mean radial polarization $\mA^\perp(t)/\bar{n}$, from BD simulations in a lattice with $\epsilon=0.6$. (a, Inset): Steady-state spatial distribution of particle number density, showing the boundary layer of thickness $\delta$ with higher number density than the bulk. $\Pe_s=1$, $\kappa^2=0.1$ were used for the simulations. (c) Comparison of boundary layer density ratios for various $\Pe_s$: $\nA/n_\infty$ from numerical simulations of the Fokker-Planck (FP) equation Eq. \eqref{eq:pdenm} in an unbounded domain and $\nA/\nB$ from BD simulations in a sparse lattice with $\epsilon =0.99$. Dashed lines indicate the analytical steady-state values of FP in an unbounded domain, Eq. \eqref{eq:blavg}.}
\label{fig:Pes1AndValidation}
\end{figure}

Since the number density of the particles cannot be measured directly on the pillar surface in discrete simulations, we consider their distribution within the boundary layer $\mathcal{A}$ as a proxy [see inset of Fig.~\ref{fig:Pes1AndValidation}(a)]. We use the same boundary layer thickness as defined in Eq.~\eqref{eq:delta} for an unbounded pillar. This is justified because the boundary layer arises from the local flux balance between the active swimming and thermal diffusion of the particles near the wall, and the influence of the bulk is minimal, except under strong confinement when accumulation layers on neighboring pillars begin to overlap. The average number density in the boundary layer is defined as
\begin{equation}
    \nA(t) = \frac{\int_\mathcal{A}  n(\rb,t)\; \mathrm{d}S}{\int_\mathcal{A} \mathrm{d}S} \approx \frac{N_\mathcal{A}(t)}{A},
    \label{eq:na}
\end{equation}
where $A=\pi \delta (\delta+2)$ is the dimensionless area of the boundary layer $\mathcal{A}$, which depends only on $\Pe_s$ and $\kappa$, and $N_\mathcal{A}$ is the instantaneous number of swimmers present in the  layer. In case of low porosity media, in which boundary layers of adjacent pillars overlap, the overlapping area is excluded from the calculations. The number density in the remaining bulk, $\mathcal{B}$, is given by
\begin{equation}
    \nB(t) = \frac{\int_\mathcal{B}  n(\rb,t)\; \mathrm{d}S}{\int_\mathcal{B} \mathrm{d}S} \approx \frac{N-N_\mathcal{A}(t)}{A_\mathcal{F} - A}.
    \label{eq:nb}
\end{equation}
Note that number conservation within the doubly-periodic cell implies $\nA$ and $\nB$ are constrained by the condition,
\begin{equation}
    \frac{\nA(t)}{\bar{n}} \frac{A}{A_\mathcal{F}} + \frac{\nB(t)}{\bar{n}}\left(1-\frac{A}{A_\mathcal{F}}\right) = 1.
    \label{eq:conservation}
\end{equation}

The evolution of $\nA$ and $\nB$ from an initially uniform suspension of number density $\bar{n}$ is shown in Fig.~\ref{fig:Pes1AndValidation}(a). As the particles accumulate near the pillar surface, $\nA$ displays an exponential growth followed by saturation. Unlike in an unbounded domain with an infinite particle reservoir, surface accumulation reduces the overall particle number in the bulk in a porous lattice, as reflected by the concomitant decrease and saturation of $\nB$ over time. The mean radial polarization of the particles within the boundary layer is obtained from BD simulations using
\begin{equation}
    {\mA^\perp(t)} = \frac{\int_\mathcal{A}  \mb(\rb,t)\cdot \hat{\rb}\; \mathrm{d}S}{\int_\mathcal{A} \mathrm{d}S} \approx \frac{1}{A}\sum_i \left(\pb_i(t) \cdot \frac{\Rb_i(t)}{|\Rb_i(t)|}\right),
    \label{eq:pol}
\end{equation}
where $\Rb_i$ and $\pb_i$ are respectively the instantaneous position and orientation of particle $i$, and the sum runs over all $N_\mathcal{A}$ particles within the accumulation layer. Here again, in Fig. \ref{fig:Pes1AndValidation}(b), we observe an exponential growth followed by saturation of the mean radial polarization, $\bar{m}_\mathcal{A}$, with the negative value indicating a net particle orientation towards the pillar. On the other hand, we find that the mean azimuthal polarization remains negligible at all times, which is expected from the symmetry of the problem.

In the limit of $\epsilon\to 1$, the suspension far from the pillar remains unaffected, so that $n_\infty = \bar{n}$. Eq.~\eqref{eq:nb} can then be rewritten as $\nB=n_\infty (1-N_\mathcal{A}/N)/(1-A/A_\mathcal{F})$. Evidently, $A/A_\mathcal{F} \to 0$ in this case. Moreover, while the number density is larger within the boundary layer, the smaller area ensures $N_\mathcal{A}/N \to 0$. Thus, in this limit, $\nB\to n_\infty$. Accordingly, the ratio $\nA/\nB$ in a porous medium serves as the analogue to $\nA/n_\infty$ defined in Eq. \eqref{eq:BLmean} for an unbounded domain. We, therefore, define the following quantities
\begin{equation}
    \eta(t) = \frac{\nA(t)}{\nB(t)},\quad \mathrm{and} \quad \mu(t) = \frac{1}{\eta(t)}\frac{{\mA^\perp(t)}}{\nB(t)}=\frac{{\mA^\perp(t)}}{\nA(t)}
\end{equation}
as measures of the number density and radial polarization, respectively, in the boundary layer. Note that $\mA/\nB$ is normalized by $\eta$ in the definition of $\mu$, {and therefore, simply measures the mean radial polarization independent of density,} following the standard convention in liquid crystalline theory \cite{Marchetti13,SaintillanShelley13}. In the limit of $\epsilon\to 1$, we expect $\eta(t)$ and $\mu(t)\eta(t)$ to approach the solution of the Fokker-Planck equation in an unbounded domain, Eq. \eqref{eq:pdenm}. We test this by comparing against the values obtained from BD simulations in a lattice of porosity $\epsilon=0.99$. Fig. \ref{fig:Pes1AndValidation}(c) shows the evolution of $\eta(t)$, with excellent agreement between the two approaches for $\Pe_s \leq 1$, the relative difference remaining below $5\%$. Although not shown here, the polarization fields also exhibit similar agreement. However, at higher $\Pe_s$, an inconsistency quickly develops between them: the relative difference in the steady-state mean number density in the boundary layer increases to $18\%$ for $\Pe_s=2$ and $55\%$ for $\Pe_s=5$. This deviation arises because of the breakdown of the $\boldsf{Q}=\mathbf{0}$ closure assumption in the Fokker-Planck description at large activity. Consistent with this, studies on transport in channels have reported significant nematic order in the near-wall region when $(\mathrm{Pe}_s/\kappa)^2 \gtrsim 10$ \cite{Ezhilan15}, which corresponds to $\Pe_s\gtrsim 1$ in our simulations. 

In the following sections, we report the results of our BD simulations in the square lattice, explaining the effects of the various control parameters viz. activity of particles, porosity, polydispersity, and flow through the medium.


\subsection{Effect of medium porosity}
\label{sec:porosity}

\begin{figure}
\centering
\includegraphics[width=\textwidth]{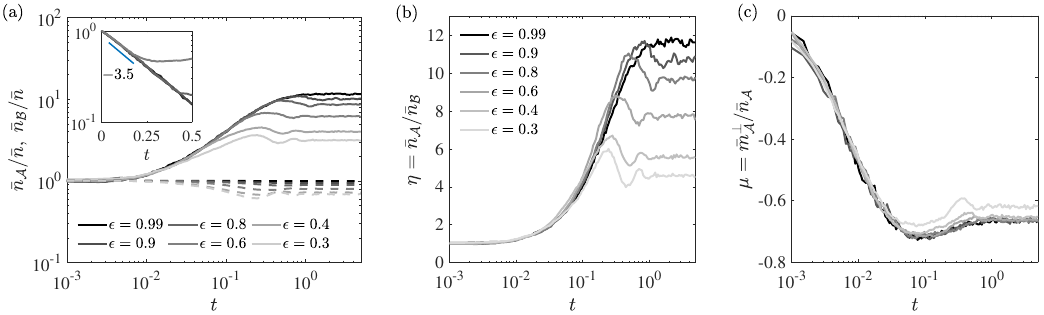} 
\caption{Time evolution of (a) the mean number densities in the boundary layer, $\bar{n}_A$ (solid lines), and in the bulk, $\bar{n}_B$ (dashed lines), normalized by the initial uniform number density, $\bar{n}$. (a, Inset): time evolution of $(\nA(t)-\eq{\nA})/(\bar{n}-\eq{\nA})$ to highlight exponential growth, with growth rate indicated by negative of the slope. (b,c) Evolution of their ratio, $\eta=\bar{n}_A/\bar{n}_B$ and of the normalized radial polarization, $\mu = \bar{m}_A/\bar{n}_A$, respectively. Values $\Pe_s=5,\; \kappa^2=0.1$ with an initial number density $\bar{n}=10^5$ were used for the BD simulations.}
\label{fig:effect_porosity}
\end{figure}

Compared to an unbounded domain, a lattice of porosity $\epsilon<1$ not only constrains the particle dynamics but also limits the total number of particles in the system. For a given initial number density $\bar{n}$, the number of active particles available for surface accumulation scales with the area of the domain, $A_\mathcal{F} \sim \epsilon/(1-\epsilon)$. On the other hand, the probability of collision of these particles with the pillar scales as $1-\epsilon$. The combined effect leads to a linear dependence of the boundary layer particle number on $\epsilon$. This trend is indeed observed in Fig.~\ref{fig:effect_porosity}(a), where the steady-state number density in the boundary layer scales roughly as $\nA/\bar{n} \sim \epsilon$. 

The reduced area of the bulk and the particle number from a decrease in $\epsilon$ also modify the bulk statistics. Assuming the boundary layer is small compared to the size of the domain, i.e. $A/A_\mathcal{F} \ll 1$ in the conservation equation Eq.~\eqref{eq:conservation}, valid except for weak activity or strong confinement ($\Pe_s\ll1,\;\epsilon \ll 1$), we find that the steady-state number density in the bulk $\nB/\bar{n}$ also decreases with a decrease in $\epsilon$, albeit more slowly than $\nA/\bar{n}$. Our simulations show that the overall effect on their ratio, $\eta\;(=\nA/\nB)$, is then a sub-linear scaling, varying empirically as $ \eta \sim \epsilon^{0.8}$ [see Fig.~\ref{fig:effect_porosity}(b)]. A reduction in the surface number density of particles with increasing confinement has also been observed in channels \cite{Ezhilan15,Ezhilan15b,Peng20}.

We now examine the transient evolution of the number densities in Fig.~\ref{fig:effect_porosity}(a). Interestingly, the early-time rate of particle accumulation on the pillar, $\dot{\bar{n}}_\mathcal{A}/\bar{n}$, is independent of the porosity. This is because the accumulation is governed by the local particle flux near the pillar, which is determined solely by $\Pe_s$ and $\kappa$ at short times, both of which are fixed in these simulations. However, under strong confinement when the boundary layers overlap (e.g., when $\epsilon=0.3$ in Fig. \ref{fig:effect_porosity}), even local interactions with the pillars become affected, leading to a slight modification in the statistics. The inset in Fig.~\ref{fig:effect_porosity}(a) shows that the growth and saturation of $\nA/\bar{n}$ has an exponential form similar to Eq.~\eqref{eq:exponentialGrowth}, 
\begin{equation}
     \frac{\nA(t)-\eq{\nA}}{\bar{n}-\eq{\nA}}=n_1 e^{-kt},
     \label{eq:naExponential}
\end{equation}
where $\eq{\nA}$ is the saturation value in the unbounded domain. {For strong activity ($\Pe_s=5$) as shown in Fig.~\ref{fig:effect_porosity}(a), a single growth rate $k\approx 3.5$ is observed for all porosities.} The porosity of the medium controls the saturation timescale; a smaller system size in the case of lower porosity reduces the average distance particles have to travel to reach the pillar, leading to faster equilibration. The identical growth rates, combined with the porosity-dependent saturation times, provides an alternate explanation for the observed variation in steady-state values of $\nA/\bar{n}$. The evolution of the bulk number density, $\nB$, follows Eq. \eqref{eq:conservation}, and therefore, indirectly depends on the porosity $\epsilon$. This is reflected in the different growth rates of $\nB$ for different values of $\epsilon$ in Fig. \ref{fig:effect_porosity}(a). Consequently, the growth rates of $\eta=\nA/\nB$ vary with $\epsilon$, slightly increasing for lower porosity, as seen in Fig. \ref{fig:effect_porosity}(b).  

Strikingly, the number densities undergo transient, damped oscillations prior to reaching their steady-state values --- a feature absent in the unbounded domain. These oscillations persist across all porosities, appearing earlier and with larger amplitudes as the porosity decreases. This behavior is attributed to the ballistic motion of particles between the pillars on short timescales \cite{Kjeldbjerg22}, which is felt more strongly in closely packed lattices (lower $\epsilon$). Consistent with this view, the oscillation frequency scales inversely with the domain length, i.e., ${a/\ell} \sim \sqrt{1-\epsilon}$ [see Fig.~\ref{fig:oscillations}(d)]. The mechanism for these oscillations relies on the formation of depletion waves when an impermeable boundary is introduced in a homogeneous suspension (see Appendix \ref{app:wave} for details). The short-time imbalance in flux at the boundary due to the ballistic accumulation of ABPs creates a localized depletion in number density in the vicinity. The depleted region propagates outwards as particles from the bulk fill the cavity. The intensity of these waves decay over time due to the diffusion of the particles. When a train of these waves originating from the surrounding pillars in the lattice wash over a given pillar, its surface number density fluctuates until it converges to a steady-state (Fig. \ref{fig:effect_porosity}). An increase in activity gives rise to a stronger depletion and therefore enhanced oscillations in the surface number density, as seen in Fig.~\ref{fig:effect_activity}(a).

The radial polarization ${\mA^\perp(t)/\bar{n}}$ also exhibits prominent oscillations before saturating when $\epsilon<1$. When normalized by the number density, the oscillations in $\mu(t)={\mA^\perp(t)}/\nA(t)$ are negligible, implying that they result from fluctuations in the number density, $\nA(t)$ [see Fig. \ref{fig:effect_porosity}(c)]. In Fig. \ref{fig:effect_porosity}(c), we see that $\mu(t)$ exhibits a peak much before $\eta(t)$ and slowly saturates to a lower value thereafter. Furthermore, $\mu(t)$ evolves independently of $\epsilon$ because particle orientations are only determined by local particle-pillar interactions. However, at low porosities (e.g., $\epsilon=0.3$), where the boundary layers overlap, particle orientations on one pillar oppose those on neighboring pillars within the overlapping region, leading to a reduction in $\mu$.

\subsection{Effect of swimmer activity}
\label{sec:activity}

\begin{figure}
\centering
\includegraphics[width=\textwidth]{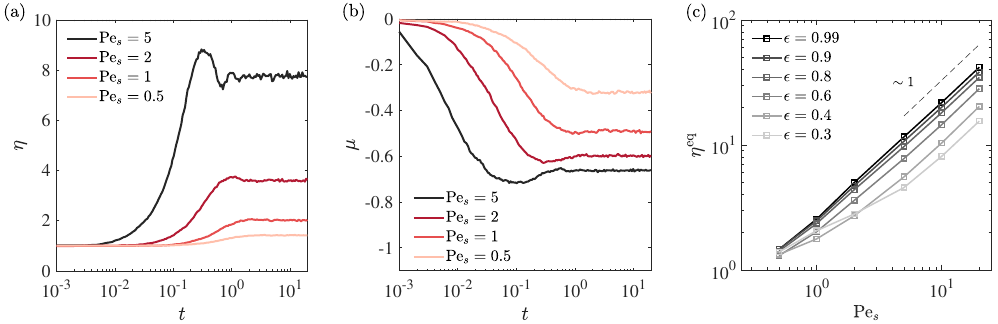} 
\caption{Time evolution of (a) surface number density and (b) radial polarization for various $\Pe_s$ in a lattice of porosity $\epsilon=0.6$. $\kappa^2=0.1$ in the BD simulations. (c) Variation of the surface number density with the $\Pe_s$ for various $\epsilon$, showing linear scaling when $\Pe_s \gg 1$.}
\label{fig:effect_activity}
\end{figure}

For fixed $\kappa$, increasing the activity (i.e., $\Pe_s$) increases the relative persistence length of the particle trajectories. Unlike a change in $\epsilon$ discussed in the previous section, varying $\Pe_s$ therefore directly alters the particle-pillar interactions and boundary layer properties. A higher $\Pe_s$ prolongs the residence time of particles on the pillar, consequently increasing the steady-state number density, $\eq{\eta}$, as seen in Fig.~\ref{fig:effect_activity}(a). As in the case of the unbounded domain, increasing $\Pe_s$ also creates sharper gradients near the surface and a reduced boundary layer thickness $\delta$.  Fig.~\ref{fig:effect_activity}(b) shows that the steady-state value of the normalized radial polarization, $\eq{\mu}$, also increases with $\Pe_s$, and appears to saturate for $\Pe_s \gg 1$.

A summary of the steady-state surface accumulation $\eq{\eta}$ with activity and porosity is shown in Fig. \ref{fig:effect_activity}(c). For strong activity ($\Pe_s \gg 1$), a lower porosity yields significantly smaller values of $\eq{\eta}$ as reasoned in the previous section. In contrast, in the weak-activity regime ($\Pe_s\ll 1$), $\eq{\eta}$ does not decrease significantly with a decrease in $\epsilon$. This is because the boundary layers in this case are thicker and, in low-porosity environments, they overlap and increase the local number density, keeping $\eq{\eta}$ roughly unaffected by $\epsilon$. One can gain insights by considering the solution in the limit of $\epsilon \to 1$. When $\Pe_s\gg 1$ in this case, the following approximate solution can be derived using the Fokker-Planck description (see Appendix \ref{app:approx}):
\begin{equation}
    \eq{\eta} \to \frac{\eq{\nA}}{n_\infty} \approx 1+\frac{0.632}{\kappa^2}\left(\frac{\Pe_s}{\sqrt{2}}-1\right), \qquad \eq{\mu} \approx \frac{-1}{\sqrt{2}}+\frac{\kappa^2}{0.632\;\Pe_s}.
    \label{eq:approx1}
\end{equation}
While $\eq\eta$ in Eq.~\eqref{eq:approx1} is close to the full solution Eq.~\eqref{eq:blavg}, its magnitude deviates significantly from the BD simulations due to the inaccurary of the closure condition used in the theoretical model, as discussed in Sec. \ref{sec:periodic}. Nonetheless, we see in Fig.~\ref{fig:effect_activity}(c) that the linear scaling with $\Pe_s$ predicted in Eq.~\eqref{eq:approx1} when $\Pe_s \gg 1$ is observed in BD simulations for all porosities. The normalized radial polarization $\eq{\mu}$, on the other hand, converges to the estimate $\eq\mu \to -1/\sqrt{2}$ when $\Pe_s \to \infty$. Surprisingly, this approximation for $\eq \mu$ is accurate here because it is inherently normalized by $\eq \eta$, which mitigates the associated error. Moreover, since we have established in the previous section that $\mu(t)$ is largely independent of $\epsilon$, we see that the large-$\Pe_s$ approximation in Eq.~\eqref{eq:approx1} compares well with BD simulations even for a lattice of porosity $\epsilon=0.6$ [Fig.~\ref{fig:effect_activity}(b)].

In addition to the steady-state values, the growth rates of number density and radial polarization on the pillar increase with $\Pe_s$ due to the higher flux of faster, more persistent particles. In the example in Fig. \ref{fig:effect_activity}(a,b), and following Eq.~\eqref{eq:naExponential}, the mean absolute number density ($\nA$) has exponential growth rates of $k=\lbrace 0.4,0.7,1.3,3.5 \rbrace$ corresponding to $\Pe_s=\lbrace0.5,1,2,5\rbrace$, respectively. The higher swim speed of particles also shortens the equilibration time of the system, as seen in the faster saturation of the number density and radial polarization for higher $\Pe_s$.


\subsection{Effect of medium polydispersity}
\label{sec:polydispersity}

\begin{figure}[t]
\centering
\setlength{\tabcolsep}{5pt}
\includegraphics[width=\textwidth]{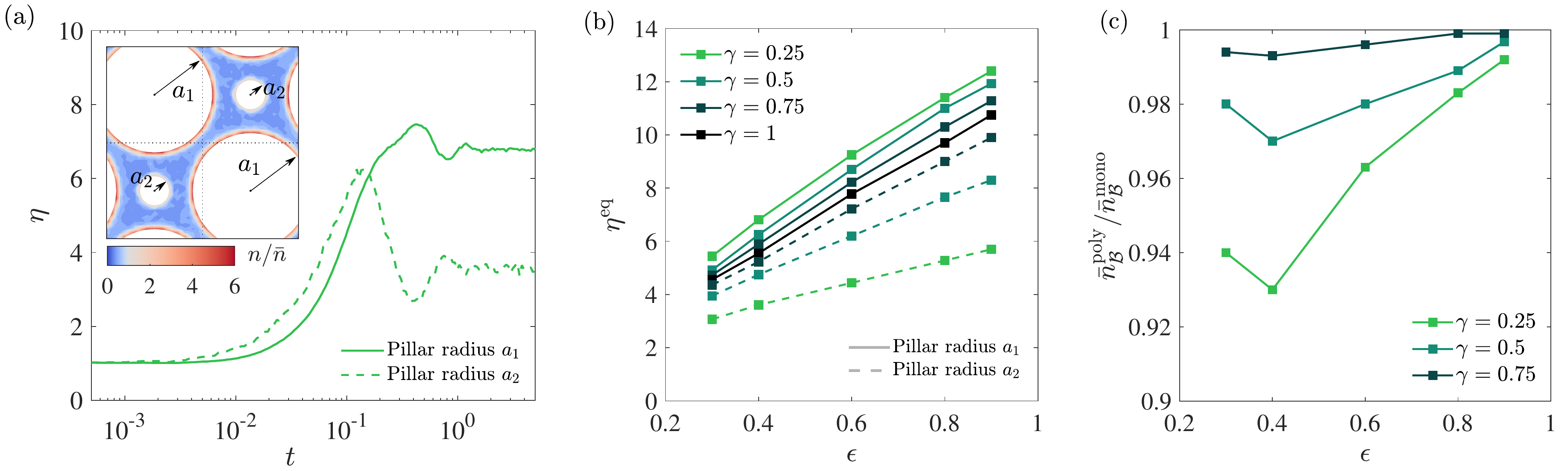}
\caption{(a) Evolution of the mean number density on pillar surfaces in a square lattice containing two pillar types of size ratio $\gamma\; (=a_2/a_1)=0.25$ and lattice porosity $\epsilon=0.4$. Inset: {steady-state} distribution of the particles in the doubly-periodic cell of the lattice. $\Pe_s=5$, $\kappa^2=0.1$ are the parameters used in BD simulations. (b) Variation of the steady-state mean number densities on the surface with $\epsilon$ for various $\gamma$. (c) Comparison of the steady-state mean number densities in the bulk between polydisperse and monodisperse lattices of identical porosity.}
\label{fig:effect_poly}
\end{figure}

Natural and engineered porous media typically comprise obstacles of varying sizes. To isolate the effect of polydispersity on the distribution of active particles, we consider the special case of a square lattice containing pillars of two distinct radii, $a_1$ and $a_2$. The doubly-periodic nature of the lattice is preserved by arranging four pillars in a diagonally-symmetric configuration within a repeating cell of side $\ell$, as shown in Fig. \ref{fig:effect_poly}(a, inset). For a given porosity $\epsilon$ and size ratio $\gamma =a_2/a_1 \leq 1$, the radii of the pillars in the cell are 
\begin{equation}
    a_1=\sqrt{\frac{1-\epsilon}{2\pi(1+\gamma^2)}}\;\ell, \qquad a_2 = \gamma a_1.
    \label{eq:pillar_sizes}
\end{equation}
One cannot, however, choose arbitrarily small values of $\epsilon$ in Eq. \eqref{eq:pillar_sizes}. Since the pillars are constrained to a square lattice and are allowed to have only two possible sizes, the permissible values of porosity have a lower bound, which depends on the size ratio $\gamma$ as 
\begin{align}
    \epsilon \geq 
    \begin{cases}\displaystyle
        1-\frac{\pi (1+\gamma^2)}{4}\qquad \textrm{if} \;\; 0 < \gamma \leq \sqrt{2}-1,\\[6pt]
        \displaystyle 1-\frac{\pi(1+\gamma^2)}{2 (1+\gamma)^2} \qquad \textrm{if} \;\; \sqrt{2}-1 < \gamma \leq 1.
    \end{cases}
\end{align}
The minimum porosity that can be achieved in this lattice is $\epsilon_\mathrm{min} = 1-\pi (1-1/\sqrt{2}) \approx 0.08$ when $\gamma = \sqrt{2}-1$. This value is much smaller than the lower bound achievable in the case of monodisperse pillars ($\epsilon_\mathrm{min} = 0.21$). The effect of confinement on the distribution of swimmers is therefore expected to be potentially more significant in the polydisperse case. The root mean square value $a_\mathrm{rms}$ of the pillar radii in Eq.~\eqref{eq:pillar_sizes} is chosen as the characteristic length scale in the problem. Upon rescaling using $a_\mathrm{rms}$, the dimensionless length of the rescaled cell is then $\ell^* = \sqrt{4\pi/(1-\epsilon)}$ and the dimensionless radii of the pillars are $a_1^*= \sqrt{2/(1+\gamma^2)}$ and $a_2^* = \gamma a_1^*$. Note that $a_1^*\geq 1$ and $a_2^*\leq 1$, so one pillar is larger and the other smaller than in the monodisperse case (unit radius). The boundary layer thicknesses corresponding to each pillar are also then rescaled as $\delta_1 = \delta a^*_1,\;\delta_2 = \delta a^*_2$, where $\delta$ is defined for a pillar of unit radius in Eq.~\eqref{eq:delta}.  

Results from a typical BD simulations performed in this dimensionless lattice are shown in Fig.~\ref{fig:effect_poly}(a). As we have seen earlier, the interaction of a particle with a pillar depends on the ratio of its run length to the pillar size, as captured by the swim P\'eclet number: a larger $\Pe_s$ implies higher accumulation rate. Here, $\Pe_s$ is defined with respect to $a_\mathrm{rms}$ rather than individual pillar radii. The local interaction is better characterized by an effective P\'eclet number for each pillar type, $\Pe_s^{(1)} = \Pe_s /a^*_1$, and $\Pe_s^{(2)} = \Pe_s /a^*_2 = \Pe_s^{(1)}/\gamma\; (\geq  \Pe_s^{(1)})$. The larger value of $\Pe_s^{(2)}$ accounts for the faster initial growth of surface number density on the smaller pillar. A striking feature in the dynamics is the presence of large fluctuations in the number density within the accumulation layers, particularly for the smaller pillar, where the means umber density peaks at nearly twice its steady-state value. This behavior is most pronounced when both $\epsilon$ and $\gamma$ are small, where the statistics are dictated by the larger pillars. Their larger perimeter accumulates more particles, thereby depleting the bulk population significantly. This depletion induces a sudden drop in number density on the surface of the smaller pillar, and a reduction in the accumulation rate on the larger ones. As the much smaller pillar does not contribute significantly to these dynamics and effectively acts as part of the bulk, the oscillations in its surface number density appear out of phase with the larger pillar. At long times, the bigger pillar has a larger mean surface number density of particles, $\eta^\mathrm{eq}$, despite its larger perimeter. In the absence of complexities from boundary layer overlap, this non-trivial feature is independent of porosity and swimmer activity, suggesting it is an intrinsic property of transport in a porous medium [see Fig.~\ref{fig:effect_poly}(b)]. By increasing the size contrast between the pillars (decreasing $\gamma$), the difference in surface accumulation becomes even more pronounced.


A porous medium can act as a filter for contaminants (bacteria and other colloidal particles). Sand filters are one such example where the bacteria aggregate on the sand particles and are retained there while the solvent medium flows out of the porous filter. How efficient a pillar matrix is at filtering out microswimmers from a fluid is a question of interest \cite{Mino18}. The effectiveness of such a filter can be estimated by comparing the ratio of the average number density of swimmers in the bulk fluid ($\nB$) and the mean number density of the whole suspension ($\overline{n}$). We had previously compared the effectiveness of the monodisperse lattice for various porosities in Section \ref{sec:porosity}. Here, we compare the effectiveness of filtering in polydisperse versus monodisperse cases in Fig. \ref{fig:effect_poly}(c). 
We see that for a given porosity in an ordered lattice, the steady-state number density of the particles in the bulk is always lower in the case of a polydisperse medium compared to monodisperse case ({$\nB^{\mathrm{poly}}/\nB^{\mathrm{mono}}\leq1$}). This is surprising given that for a fixed $\epsilon$, the available boundary perimeter for particles to aggregate on is less in the polydisperse case by a factor of $(1+\gamma)/\sqrt{2(1+\gamma^2)}$. {The ratio $\nB^{\mathrm{poly}}/\nB^{\mathrm{mono}}$ is found to reach a minimum at $\epsilon\approx 0.4$.}

\section{Effect of flow through the medium}
\label{sec:flow}

Microswimmers such as bacteria are often found in saturated porous environments both in the their natural habitats as well in industrial settings where they encounter fluid flows that can affect their surface colonization. The long-time distribution of passive colloids in a doubly-periodic lattice has been well-studied using Taylor Dispersion theory \cite{Brenner93}. More recently, the theory has been extended to account for particle self-propulsion \cite{AlonsoMatilla19}. 
Here, we focus our attention on the role of an external flow on the distribution of active particles in a two-dimensional lattice. We impose the flow in the square lattice by applying a pressure jump across the unit cell. To keep the analysis simple, we consider the pressure jump only one direction, say, the $x$-direction i.e. between the left and right walls of the unit cell. The resulting Stokes flow is computed to machine-precision using the Method of Fundamental Solutions \cite{Barnett15,Barnett18} (see Appendix \ref{app:BIM}), and the typical flow field is shown in Fig. \ref{fig:flow_field}. The strength of transport by flow through the medium is determined by the flow P\'eclet number,
\begin{equation}
    \Pe_f = \frac{u_\infty}{a D_R},
\end{equation}
where the characteristic velocity, $u_\infty$, is defined in terms of the flux crossing a unit cell, and is therefore, an estimate of the mean velocity of the fluid in the cell; for a fixed $u_\infty$, the mass flux of fluid through the lattice is the same regardless of porosity.

We model the microswimmers as elongated point particles, a good approximation for many rod-shaped bacteria and catalytic colloidal rods. In a background shear flow, slender particles exhibit flow-aligning behavior, with their dynamics following Jeffery's equation \cite{Jeffery1922,Bretherton62,DoiEdwards}. In the dilute limit, the Langevin equations governing the particle dynamics modify to
\begin{subequations}
    \begin{align}
    \dot{\Rb}(t) & = \Pe_s\; \pb(t) + \Pe_f\; \ub(\Rb) + \sqrt{2 \kappa^2}\, \boldsymbol{\xi}_T(t) \label{eq:Jeff1}, \\ \dot{\pb}(t) & = \Pe_f\; \pb(t) \times (\bt{G}(\Rb) \cdot\pb(t)) \times \pb(t) + \sqrt{2}\, \boldsymbol{\xi}_R(t) \times \pb(t).
    \label{eq:Jeff2}
\end{align}
\end{subequations}
Here $\bt{G}(\rb)=\bnabla \ub({\rb})$ is the dimensionless velocity gradient tensor, where $\ub({\rb})$ is the velocity field in the rescaled cell, normalized using $u_\infty$ (see Appendix \ref{app:BIM}). Unlike the Poiseuille flow in a channel considered in many previous studies, the present flow field has a fully two-dimensional structure, with components of $\bt{G}$ varying in both the streamwise and cross-streamwise directions (see Fig.~\ref{fig:flow_field}). In the periodic single pillar case, the vorticity field is mirror-symmetric about the $x$-axis, with a single sign on either side. 

\subsection{Transient dynamics}
\label{sec:transient}

\begin{figure}
    \centering
        \ 
        \includegraphics[width=\textwidth]{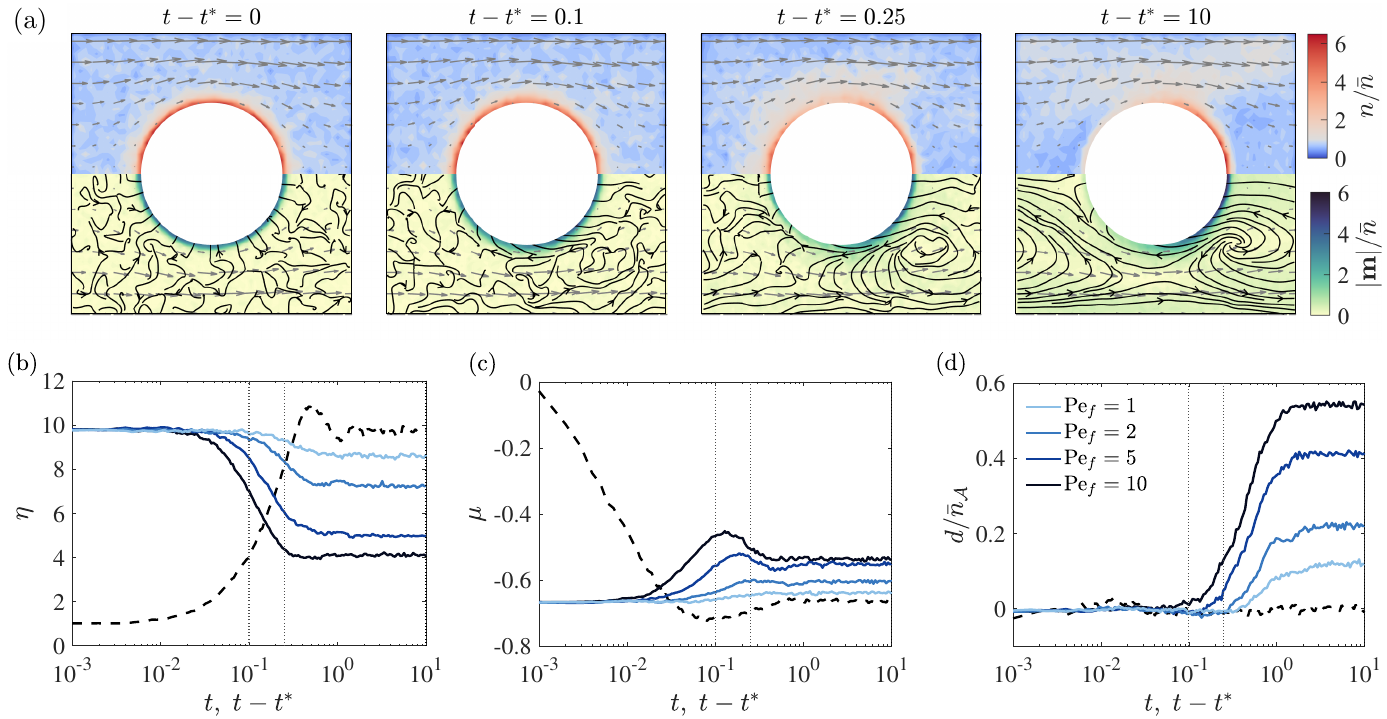}
    \caption{Time evolution of the distribution of particles in the presence of background flow. The flow is imposed from left to right at some time $t=t^*$, after the system reaches a steady state. (a) Instances of evolution of the number density $n(\rb,t)$ and polarization $\mb(\rb,t)$ fields due to the fluid flow (shown in gray arrows) in the case of $\epsilon=0.8$, $\Pe_s=5$, and $\Pe_f=5$. The fields are mirror-symmetric about the $x$-axis. Also shown are the evolution of the boundary layer-averaged quantities, including (b) number density ratio, $\eta$ (c) radial polarization, $\mu$, and (d) moment of the particle distribution on the pillar (along the $x-$direction, by symmetry), $d=\bar{\bm{d}}_\mathcal{A}\cdot \hat{\xb}$, for various $\Pe_f$ while other parameters are kept the same as in (a). The dashed lines show the evolution of these quantities before flow is initiated, and the dotted lines indicate the instances shown in (a) after flow is initiated.}
    \label{fig:flow_distr}
\end{figure}

To isolate and understand the effect of flow on the particle distribution, we let the system evolve without any flow until a steady state is reached, after which we initiate flow at some $t=t^*$. For illustration, we consider the case with $\epsilon=0.8,\;\Pe_s=5,$ and $\Pe_f=5$ shown in Fig.~\ref{fig:flow_distr}. As before, the results of BD simulations of Eqs.~\eqref{eq:Jeff1}-\eqref{eq:Jeff2} are interpreted using the coarse-grained Fokker-Planck description. Figure~\ref{fig:flow_distr} shows the evolution of the number density and polarization fields in the lattice after the flow is introduced. Note that flow breaks the symmetry of the fields, and therefore, to quantify each, we introduce two additional measures, namely ($i$) the dipolar moment of the surface number density, $\qb(t)=\int n_s(\nb,t)\, \nb\; \mathrm{d}S$, with $n_s$ and $\nb$ being the local number density on the surface and outward surface normal respectively, and ($ii$) the net upstream swimming flux, $\mathcal{U}(t)$, defined later. The boundary layer-average of the moment on the circular pillar at any time $t$ is then
\begin{equation}
        \qbar(t) = \frac{1}{A} \int_\mathcal{A} \frac{n(\rb,t) \rb}{r} \; \mathrm{d}S \,\approx\, \frac{1}{A}\sum_i \frac{\Rb_i}{|\Rb_i|},
\end{equation}
where $\Rb_i$ is the position vector of a particle $i$ and the sum runs over all $N_\mathcal{A}$ particles within the boundary layer. Assuming that the particle distribution around the pillar does not vary significantly within the boundary layer (i.e., $\qbar \approx \qb$), then $\qbar$ relates to the swim force on the pillar via \cite{YanBrady15}
\begin{equation}
    \boldsymbol{F}(t) = -k_BT \qbar(t).
    \label{eq:swimForce}
\end{equation}

At steady-state prior to flow initiation, the distribution of particles on the pillar surface is uniform and aligned normal to it. Here, $\qbar =\mathbf{0}$, by symmetry, and therefore, there is no net swim force on the pillar. Upon initiation of the flow, at short times $(t - t^*) < 0.05$ --- marked by no significant change in the surface number density, $\eta$ (Fig.~\ref{fig:flow_distr}(b)) --- the shear near the surface of the pillar rotates the particles against the direction of flow. The radial polarization $\mu$ decreases accordingly within the boundary layer (Fig.~\ref{fig:flow_distr}(c)). The distribution of the particles around the pillar remains uniform during this short period ($d\approx 0$) as the advective transport is weak near the surface.

At longer times ($0.05 \lesssim (t - t^*) < 1 $), $\eta$ decreases because the particles reoriented away from the surface by shear swim into the bulk where they are advected by a stronger flow. $\mu$ attains a minimum during this time, as the strain aligns many particles within the boundary layer tangential to the surface. The strong vorticity competes with this strain to reorient particles towards the cross-stream direction (see Fig.~\ref{fig:flow_field}). From the slightly weaker strain regions towards the rear of the pillar, particles are reoriented and swim in the cross-stream direction, where they are simultaneously advected and rotated by the flow towards a downstream-aligned orientation in the center of the fluid channel. Crossing the periodic domain, the particles experience the same vorticity in the front of the pillar, which reorients them again, this time towards the front of the pillar. For the relatively moderate values of $\Pe_f$ considered here, this effect results in swirling structures in the polarization field; the two swirls have opposing sense of rotation, reflecting the local vorticity (see Fig.~\ref{fig:flow_distr}(a) panel 3). Moreover, the polarization field develops regions of upstream swimming at the wake of the pillar, which increases the radial polarization $\mu$, the local number density $n$, as well as the dipolar moment, $d$.

At very long times $(t - t^*) \gg 1 $, the particles reach a new steady state distribution within the domain. The boundary-averaged quantities $\eta$ and $\mu$ saturate to their new steady-state values, $\eta^\mathrm{eq}$ and $\mu^\mathrm{eq}$, lower than their values in the absence of flow. In fact, $\eta^\mathrm{eq}$ decreases with increasing $\Pe_f$ and asymptotes towards 1 as $\Pe_f \to \infty$. The new steady-state dipolar moment, however, is non-zero due to the accumulation of particles at the leeward side of the pillar by upstream swimming. The presence of positive rheotaxis in the wake is in accordance with the experimental observations \cite{Secchi20,Mino18}. In fact, distinct upstream (around the pillar) and downstream (top and bottom cell boundaries) swimming regions form at steady state. The swirling structures in the polarization field tighten to form spiral defects at long times, with a pair of counterbalancing hyperbolic defects; together, they act as a topological connection between the upstream- and downstream-swimming regions. The outward spiral defects are regions in the domain from which particles swim away, and therefore act as a local depletion zone for the particle number density (see Fig. \ref{fig:flow_distr}(a) panel 4). We analyze these defects in more detail next.  

\subsection{Flow-induced topological defects and swimming transition}
\label{sec:defect}

\begin{figure}
    \centering
    \includegraphics[width=\textwidth]{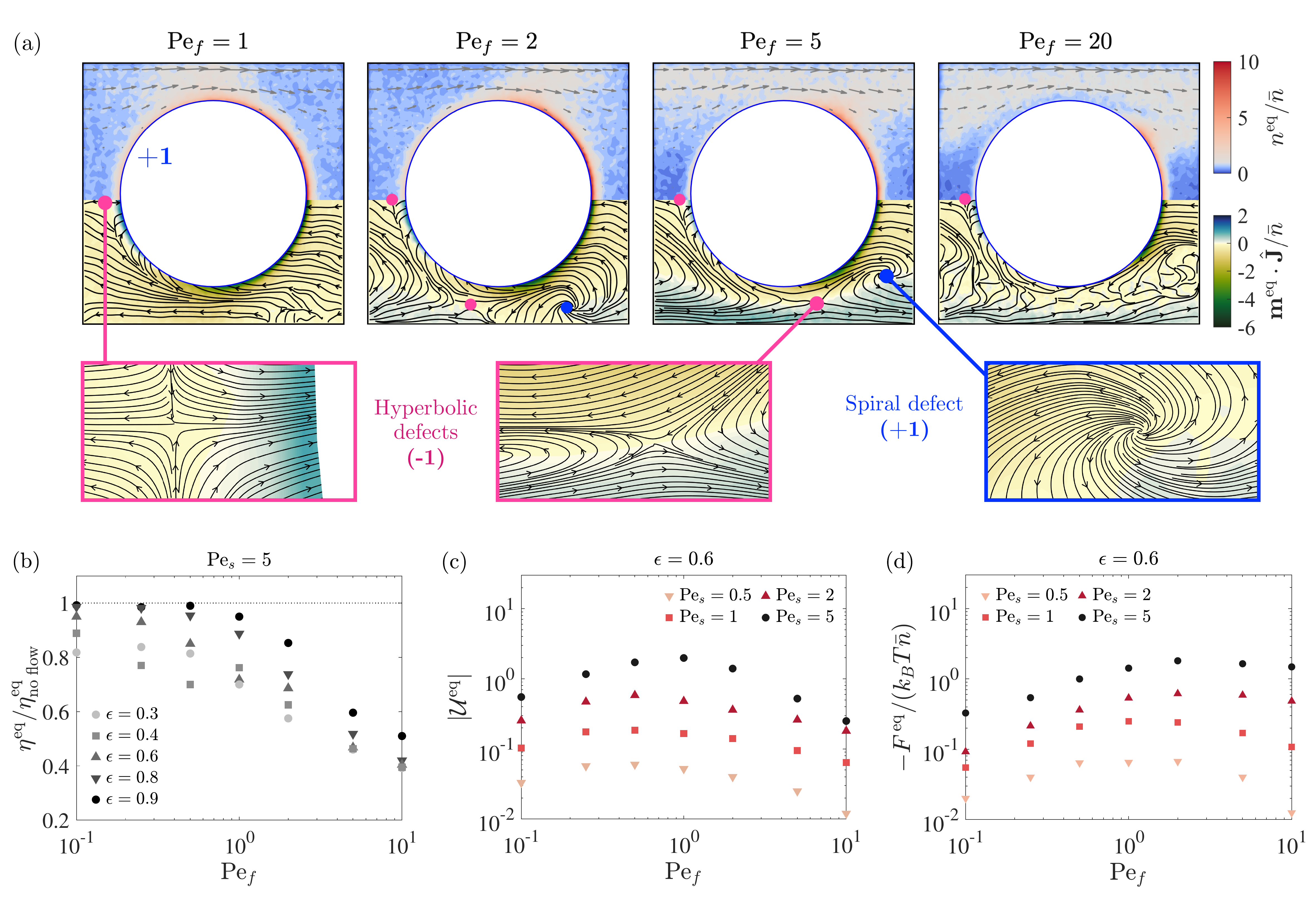}
    \caption{Steady-state statistics in a square lattice in the presence of flow: (a) The steady-state number density $n^\mathrm{eq}(\rb,t)$ and polarization $\mb^\mathrm{eq}(\rb,t)$ fields when $\epsilon=0.6$ and $\Pe_s=5,\; \kappa^2=0.1$ for various flow strengths, $\Pe_f$. The pillar surface has a +1 topological charge. The creation of defect pairs, viz. hyperbolic ($-1$) and spiral (+1), at moderate $\Pe_f$ marks the transition from globally upstream swimming ($\smash{\mb^\mathrm{eq}\cdot\hat\Jb<0}$) to distinct upstream/downstream regions. (b) Reduction in the particle number density on the surface with $\Pe_f$ for all porosities, $\epsilon$. (c) Upstream swimming flux of the particles in the lattice and (d) their swim force on the pillar surface ($F=\Fb\cdot\hat{\xb}$) as functions of $\Pe_f$ for various $\Pe_s$.}
    \label{fig:flowSteadyStateDefect}
\end{figure}

Fig.~\ref{fig:flowSteadyStateDefect}(a) shows the steady-state distribution of  particles of $\Pe_s=5$ in a square lattice of porosity $\epsilon=0.6$ for various $\Pe_f$. For weak flow strength, $\Pe_f \leq 1$, the upstream polarization spans the entire domain at steady state, whereas in the limit of strong flow, $\Pe_f \geq 20$, two distinct regions of upstream and downstream polarization can be identified. While a similar transition is also observed in channel flow \cite{Ezhilan15,Peng20}, here we find that the intermediate values of $\Pe_f \sim 1-20$ show well-defined point defects in the polarization field, absent in the channel geometry. We characterize this transition in what follows using the net upstream swimming flux of the particles as a macroscopic measure, defined at any time $t$ as 
\begin{equation}
    \mathcal{U}(t) = -\Pe_s\,\frac{\bar{\mb}(t)}{\bar n}\cdot \hat\Jb\; \approx -\Pe_s\frac{1}{N} \sum_{i=1}^N \pb_i(t) \cdot \hat\Jb,
    \label{eq:upstreamSwimFlux}
\end{equation}
where $\bar{n}$ and $\bar{\mb}$ are the area-averaged number density and polarization of the particles in the entire domain $\mathcal{F}$ of the doubly-periodic cell {(Eq.~\eqref{eq:meanDensityPolarization})}  and $\hat{\Jb}=\Jb/|\Jb|$ is the direction of the net fluid flux $\Jb$ through the cell. In the case of a square lattice considered here, with pressure-drop along the $+x$-axis, $\hat \Jb=\hat{\boldsymbol{x}}$. A positive (negative) value of $\mathcal{U}$ indicates a  net upstream (downstream) swimming flux.

As shown in previous sections, in the absence of flow, the steady-state polarization at the pillar surface is radial by symmetry; the particles at the surface are, on average, oriented towards the center of the circular pillar. Thus, the pillar introduces a net topological charge of $+1$ in the polarization field due to the normal ``anchoring" of the particles around it, and the balance negative charge ($-1$) is distributed across the unit cell outside the boundary layer (Fig.~\ref{fig:flow_distr}(a), panel 1). There is no net upstream swimming flux by symmetry, $\mathcal{U}^\mathrm{eq}=0$. The presence of a background flow modifies the particle distribution around the pillar by reorienting the microswimmers. Nevertheless, the no-flux condition requires that the polarization field maintain a non-zero radial component into the pillar everywhere on the surface, ensuring that the pillar remains 
a disclination of topological charge of $+1$. For weak flow ($\Pe_f \leq 1$ in Fig.~\ref{fig:flowSteadyStateDefect}), the shear-induced rotation of particles near the pillar, together with their relatively strong activity, enables global upstream polarization in the domain at long times, which focuses the $-1$ charge to a point defect at the front of the pillar. By increasing $\Pe_f$ within this regime, the steady-state orientation of particles move closer towards the $-x$-direction, increasing the upstream flux, $\mathcal{U}^\mathrm{eq}=\mathcal{U}(t\to\infty)$; see Fig. \ref{fig:flowSteadyStateDefect}(c).

Interestingly, with further increase in $\Pe_f$, two additional pairs of defects nucleate in the bulk: outward spiral defects of $+1$ charge and hyperbolic defects of $-1$ charge. Figure~\ref{fig:flowSteadyStateDefect}(a, panels 2--3) show one pair of defects; the other pair is mirror symmetric. The defects break the local order of upstream or downstream polarization, causing a reduction in $\mathcal{U}^\mathrm{eq}$ with increasing $\Pe_f$ (see Fig.~\ref{fig:flowSteadyStateDefect}(c)). The transient evolution of this regime was discussed in Sec.~\ref{sec:transient}, where strong shear gradients near the pillar were identified as the cause of these defects. Importantly, the shear gradients in the lattice are genuinely two-dimensional (see Fig.~\ref{fig:flow_field}), so the local orientation field is allowed to wind around a point and form topological defects; in a channel geometry, shear varies along only one direction, which makes winding topologically impossible. These (stationary) defects moving downstream on increasing the flow strength, $\Pe_f$, is indicative of their kinematic origin. The defects are also observed for other values of $\Pe_s$ where the boundary layers on pillars do not overlap. Unlike the topological defects commonly studied in active matter, where orientational defects are energetically favorable states of interacting particles, here, they originate purely from the kinematics of active particles in complex shear flow. 

For large $\Pe_f$ ($\geq 20$ in Fig.~\ref{fig:flowSteadyStateDefect}), the regions of strong shear gradients rapidly decorrelate the orientation of nearby particles in the region, thereby suppressing the spiral structure they otherwise develop downstream. The defects slowly dissolve with increasing $\Pe_f$, leaving behind distinct regions of upstream and downstream swimming, separated by a disordered region of low polarization, akin to the high-shear trapping regime in a channel geometry.

We examine the steady-state swim force $\Fb$ on the pillar (Eq.~\eqref{eq:swimForce}), resulting from the non-uniform distribution of particles on the surface in the presence of flow. The mirror symmetry of the distribution about the $x$-axis implies $\Fb=F\,\hat{\xb}$. Upstream polarization drives accumulation in the pillar's wake, resulting in a net swim force $F<0$. At small $\Pe_f$, advection reinforces this wake accumulation, thereby increasing $|F|$. At large $\Pe_f$, however, advection depletes particles from the surface while upstream swimming weakens, both contributing to a reduction in $|F|$. The crossover between these two regimes produces a peak in $|F|$ as a function of $\Pe_f$ (see Fig.~\ref{fig:flowSteadyStateDefect}(d)).

\begin{figure}
    \centering
    \includegraphics[width=\textwidth]{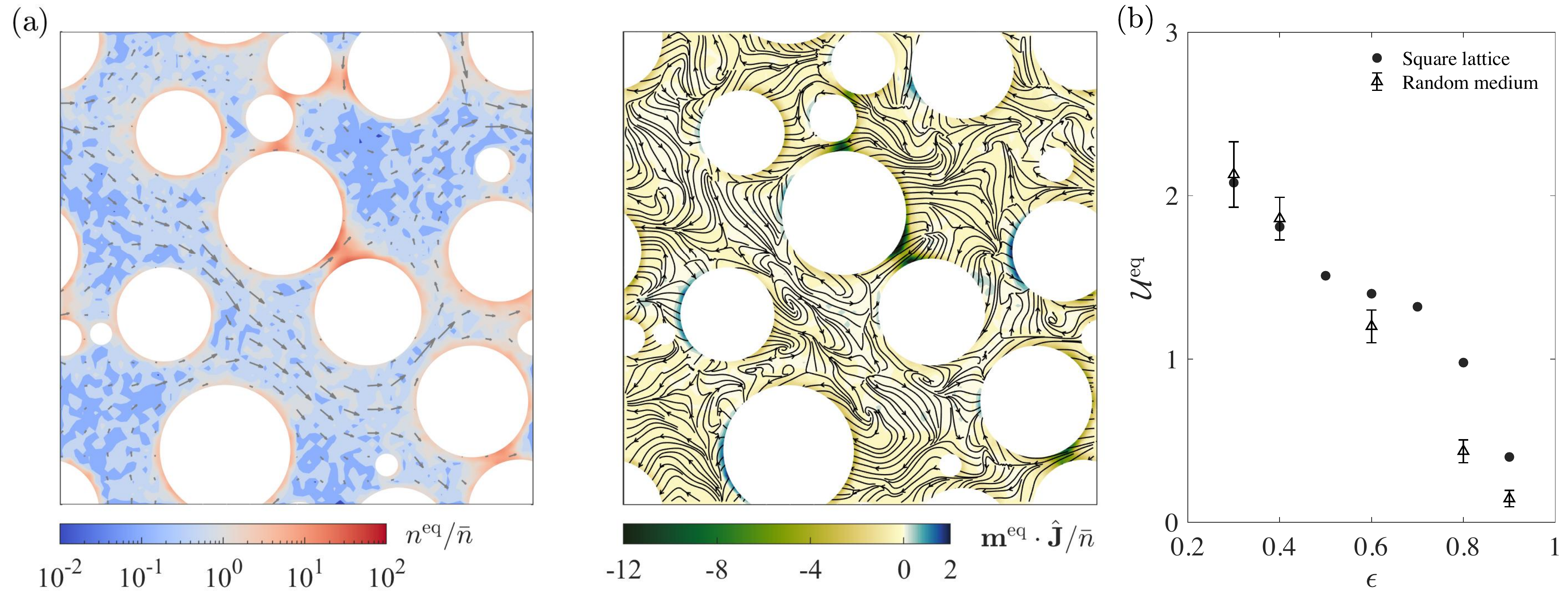}
    \caption{(a) The steady-state number density $n^\mathrm{eq}(\rb,t)$ and polarization $\mb^\mathrm{eq}(\rb,t)$ fields when $\Pe_s=5$, $\Pe_f=2$ in a random lattice of porosity, $\epsilon=0.6$ with net fluid flux along $\hat\Jb=0.993 \,\hat{\bm{x}} - 0.118\,\hat{\bm{y}}$. The magnitude of polarization in the direction of the net flow is shown to highlight the upstream/downstream swimming of the particles. (b) Comparison of the upstream swimming flux between the monodisperse square lattice and various realizations of a random medium for $\Pe_s=5$ and $\Pe_f=2$. In all cases, $\kappa^2=0.1$.}
    \label{fig:randomMedia}
\end{figure}

The positive rheotaxis at the wake locally increases the number density at the rear of the pillar. Nonetheless, as discussed earlier in Sec.~\ref{sec:transient}, the advective flux, being much stronger, decreases the mean surface number density, $\eta$, with increasing $\Pe_f$, irrespective of porosity of the medium, $\epsilon$ (see Fig.~\ref{fig:flowSteadyStateDefect}(b)). The decrease is non-monotonic due to the non-trivial active transport from the complex polarization field. For a lower $\epsilon$, both the velocity and shear at the throat are higher for a given $\Pe_f$, which enhances particle depletion from the surface, resulting in a larger reduction in $\eta$. Varying the porosity also affects the transport in many ways. For one, a lower porosity exposes more particles to the near-wall region where upstream polarization is strong, thereby enhancing the net upstream swimming flux, $\mathcal{U}^\mathrm{eq}$ (see Fig.~\ref{fig:randomMedia}(b)). Secondly, the defects in the polarization field do not persist for higher $\Pe_f$ when the porosity is lower.

The key phenomena identified in the preceding sections for the square lattice are qualitatively reproduced in a random lattice, including higher number densities on bigger pillars, wake-accumulation through upstream swimming, downstream swimming in faster channels, and spiral and hyperbolic defects in the polarization field (see Fig.~\ref{fig:randomMedia}(a)); the defects are, therefore not merely a special case due to the periodicity of the lattice. Similar to the square lattice, $\mathcal{U}^\mathrm{eq}$ in a random medium also increases with a decrease in $\epsilon$. However, $\mathcal{U}^\mathrm{eq}$ is typically higher for a square lattice because the alignment of pillars promotes upstream swimming (see Fig.~\ref{fig:randomMedia}(b)).

\section{Concluding remarks}
\label{sec:conclusions}

Surface accumulation is a characteristic property of out-of-equilibrium systems. In this work, we investigated the distribution of active particles in a two-dimensional lattice of pillars, characterizing both their temporal evolution and steady-state distribution in the medium. As with a single pillar in an unbounded domain, the surface accumulation from an initially homogeneous state proceeds through an exponential growth and saturation of the particle number density on the surface, with rates increasing with $\Pe_s$. The relaxation to this new steady state is characterized by oscillations in the surface number density due to the impact of propagating depletion waves originating from the surface of the surrounding pillars at short times. Wave formation is an active phenomenon rooted in the ballistic nature of the particles \cite{Geyer18,Dulaney20}, and these density waves, in conjunction with diffusive processes, set the timescale over which the system relaxes to its global steady state. 

The migration of passive colloidal particles in the presence of a shear flow is a well-studied phenomenon \cite{Leighton87}. This classical result continues to shape current descriptions of transport in complex flows. In active suspensions, both high-shear and low-shear trapping regimes have been reported in experiments of motile bacterial suspensions in channel flows \cite{Rusconi14,Barry15}. While shear-induced migration in passive suspensions result from interactions between colloids, in active suspensions, on the other hand, where orientational dynamics plays an important role, even non-interacting particle models have been shown to be sufficient to explain the phenomena \cite{Ezhilan15,Vennamneni20}. We used the non-interacting ABP model in our discrete particle simulations to show the emergence of a swimming transition in lattices — a global upstream polarization at small $\Pe_f$ and coexisting up- and downstream polarization at higher $\Pe_f$. Upstream swimming was found to be the cause for preferential accumulation of bacteria in the wake of the pillar \cite{Secchi20,Mino18}. Importantly, we found that the transition involves the nucleation of topological defects in the polarization field of the particles, followed by their dissolution at high $\Pe_f$. Topological defects in polar active fluids, including spiral and hyperbolic defects, remain an active area of research in active matter, motivated in part by their experimental observation in motility assays and suspensions of colloidal rollers \cite{Shankar22,Angheluta25}. While the defects identified here have a similar signature, their physical origin is fundamentally different. Conventional defects in active polar fluids arise from instabilities driven by the two-way coupling between particle orientation and the active stresses generated be the particles  \cite{Simha02,Kruse04}. By contrast, the defects reported here are stationary structures that emerge from the kinematic response of non-interacting self-propelled particles to a complex shear flow, with no feedback of particle orientation onto the flow. A genuinely two-dimensional shear profile is essential for their formation, with $\Pe_f$ serving as the control parameter for defect nucleation.

Here, we explored one of the simplest settings for active transport in porous media: a square lattice of obstacles. The periodic obstacle array creates a fully two-dimensional flow field and has historically served as a model porous medium in transport studies \cite{Sangani82,Brenner93}. This geometry offered a controlled setting in which to vary particle self-propulsion, medium porosity, and externally imposed flow. The present model, however, has several important limitations. We assumed the particles to be point-sized thereby omitting the effects of steric interactions between particle and the wall surfaces. In reality, steric interactions prevent certain particle configurations to occur at the wall, thus creating a significant reduction in the estimated probability density, which is not resolved in our simulations \cite{Chen21,Ezhilan15}. Furthermore, particle hydrodynamic fields and the resulting inter-particle and particle-wall interactions were ignored in our computations. Their effect on particle trajectories are found to be non-negligible in both experiments and numerical simulations \cite{Spagnolie15,Sipos15}. Steric and hydrodynamic interactions are particularly important in denser suspensions where they give rise to instabilities and transitions to chaos \cite{Wioloand13,Saintillan07,Alert22}. Finally, while the simple two-dimensional lattice structure allowed us to study the role of confinement and shear gradients, future studies extending to disordered media and three dimensions, while more challenging, would likely produce more complex shear flow and particle dynamics that better describe natural environments.

\section*{Acknowledgments}
D.S. acknowledges support from NSF Grant 1934199. A.V. thanks the Alexander von Humboldt Foundation for support during the final stages of this project. The authors are grateful to Pallabi Das (MPI-PKS) and Suganthan Senthilkumar (MPI-PKS) for useful discussions. 

\appendix

\section{On the propagation of the depletion front}
\label{app:wave}

Active suspensions of self-propelled particles exhibit wave-like behavior on time scales shorter than their reorientation time ($t\lesssim 1$ in dimensionless variables), eventually exhibiting diffusive behavior on longer time scales \cite{Dulaney20}. Following Sec. \ref{sec:single_unbounded}, we discuss the time evolution of the number density $n$ and the polarization $\mb$ in an initially isotropic suspension containing a singe isolated pillar of unit radius. In the absence of any external flow, the system has radial symmetry and so, these mean fields are a function of $r$ alone; the polarization field, then, only has a radial component, $\mb(r,t)=m(r,t) \hat{\rb}$. The coupled system Eqs. \eqref{eq:pdenm}--\eqref{eq:fluxnm} is solved for $n(r,t)$ and $m(r,t)$ using finite differences, with an implicit adaptive time-stepping to handle the numerical stiffness. The spatial grid outside the pillar is stretched by performing the change of variables: $r \to \rho = \exp(-\beta(r-1))$, where the user-defined scaling parameter $\beta > 0$ is selected to capture the sharp gradients inside the accumulation layer; $\beta=3$ was shown to produce results that agree well with the Brownian Dynamics simulations. 

\begin{figure}
    \centering
    \includegraphics[width=\textwidth]{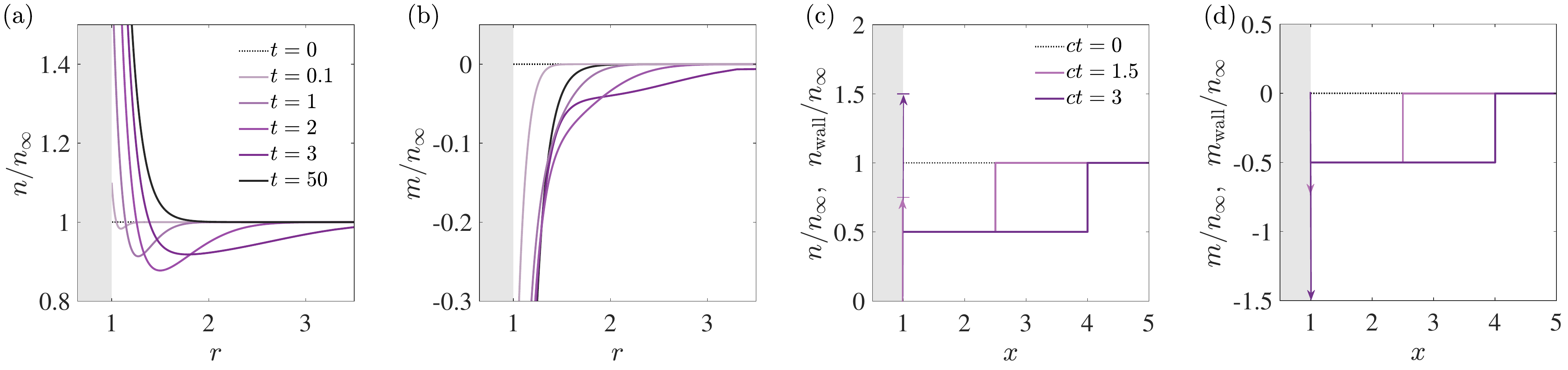}
    \caption{Numerical solution for the distribution of (a) the particle number density, $n(r,t)$, and (b) the radial polarization, $m(r,t)$, near the surface of a circular pillar of unit radius in an unbounded domain, at various instants of time, $t$, for the case of $\Pe_s=1,\;\kappa^2=0.1$. (c,d) Time evolution of $n(x,t)$ and $m(x,t)$ in a 1D domain in the absence of translational diffusion ($\kappa\to 0$), given by the analytical expressions in Eq. \eqref{eq:app_n} and Eq. \eqref{eq:app_m}, respectively. At $t=0$, the suspension is uniform and isotropic ($n=n_\infty, \; m=0$).}
    \label{fig:num_wave}
\end{figure}

The numerical solutions for  $n(r,t)$ and $m(r,t)$ obtained in this manner are shown in Fig. \ref{fig:num_wave}(a,b). The number density near the pillar increases over time, as particles oriented normal to the surface collide and remain there until rotational diffusion reorients them away. For an initially-uniform suspension of density $n_\infty$, the short-time influx of particles from the bulk to the pillar surface remains steady while the outflux is reduced due to the preferential orientation of the particles toward the pillar. This imbalance in flux, together with the ballistic nature of particles, creates a local depletion in particle density ($n/n_\infty<1$) slightly away from the pillar surface, as shown in  Fig. \ref{fig:num_wave}(a). The continued flux from the bulk progressively displaces the depleted region farther into the bulk. Due to the ballistic motion of the active particles, flux imbalances propagate as dispersive waves, rather than spreading purely diffusively as in passive Brownian systems.

\subsection{\texorpdfstring{Solution in the limit $\kappa \to 0$}{Solution in the limit kappa → 0}}
\label{app:wave_analytical}

 One can gain insights into this mechanism by studying a 1D system where the particles have only two possible orientations viz. along the positive or the negative $x$-axis. In the absence of translational diffusion, i.e., when $\kappa \to 0$, Eqs. \eqref{eq:pdenm}--\eqref{eq:fluxnm} reduce to the wave equation for the number density, $\partial^2_t n(r,t) =c^2 \partial_x^2 n(r,t)$. Here, $c=\Pe_s$ is the wave speed in 1D. For a given initial distribution $n(x,0)=f(x)$, the subsequent evolution is given by  
\begin{equation}
    n(x,t) = \int_{-\infty}^\infty G^{(1)}(x-x',t) f(x')\; \mathrm{d}x',
    \label{eq:app_n_temp}
\end{equation}
where $G^{(1)}(x,t) = \delta(ct-|x-x'|)$ is the 1D Green's function of the wave equation, with $\delta(x)$ being the Kronecker delta function. If we consider a wall at $x=1$ and particles in the domain $x>1$, an initially uniform distribution is given by $f(x)=n_\infty \Theta(x-1)$, with $\Theta(x)$ being the Heaviside step function. Substituting in \eqref{eq:app_n_temp} and evaluating, we get
\begin{equation}
    n(x,t)=\frac{n_\infty}{2}\left[\Theta(x-1-ct)+\Theta(x-1+ct)\right] \qquad \mathrm{for} \;\; x>1.
    \label{eq:app_n}
\end{equation}
The solution in $x < 1$ is unphysical and hence ignored. In the absence of diffusion, all the particles that reach the wall by swimming end up condensing there. The rate of increase of the particle number $n_{\mathrm{wall}}(t)$ at the wall thus balances the swimming flux towards it, from which
\begin{equation}
    n_\mathrm{wall}(t) = \frac{n_\infty}{2} c t.
\end{equation}
The evolution of $n(x,t)$ is shown in Fig. \ref{fig:num_wave}(c), where we indeed observe the formation of  a depletion region that propagates outwards but does not decrease in magnitude over time as there is no diffusion. The corresponding polarization field can be found using Eq. \eqref{eq:pdenm}, which states that
\begin{align}
    \pard{m}{t} =- c\pard{n}{x}= -\frac{n_\infty}{2} c \left[\delta(x-1-ct)+\delta(x-1+ct)\right].
\end{align}
Performing the integral, and noting that the suspension is initially unpolarized, i.e, $m(x,0)=0$, we obtain
\begin{equation}
    m(x,t) = -\frac{n_\infty}{2}\Theta(ct-|x-1|)  \qquad \mathrm{for} \;\; x>1.
    \label{eq:app_m}
\end{equation}
All the particles condensed at the wall must point towards the wall in order to remain there, and therefore the polarization at the surface is
\begin{equation}
    m_\mathrm{wall}(t)=-n_\mathrm{wall}(t)=-\frac{n_\infty}{2}ct.
\end{equation}
The mechanism for the depletion wave is quite simple. In the bulk far away from the wall, isotropy of the suspension implies equal fluxes of particles swimming the positive and negative $x$-directions. Close to the wall, however, particles pointing in the positive $x$-direction are not replenished due to the no-flux condition, giving rise to the depletion, which grows in size with the motion of the particles swimming away from the wall. Simultaneously, particles swimming towards the wall condense there, giving rise to particle accumulation at the wall surface. Inside the depletion region, only particles pointing towards the wall remain, hence the negative polarization. In the presence of diffusion, both translational and rotational diffusion tend to smooth gradients in the near-wall region, giving rise to the profiles of Fig.~\ref{fig:num_wave}(a,b).


A similar analysis can be carried out for the 2D radially symmetric case, where one obtains the telegraph equation $\partial^2_t n(r,t) + \partial_t n(r,t)=c^2 \nabla^2 n(r,t)$ when $\kappa \to 0$. The wave speed in this case is $c=\Pe_s/\sqrt{2}$  with the Green's function $G^{(2)}(r,t)=2\Theta[(t-r)]e^{-t/2} \cosh(\sqrt{t^2-r^2}/2)/\sqrt{t^2-r^2}$ \cite{Dulaney20}. A complicated series solution can be obtained for $n(r,t)$ in this case, which is omitted here for brevity. The qualitative behavior of the depletion front is unchanged from the 1D case.

\subsection{Features of the oscillations}

\begin{figure}[t]
\centering
\includegraphics[width=0.95\textwidth]{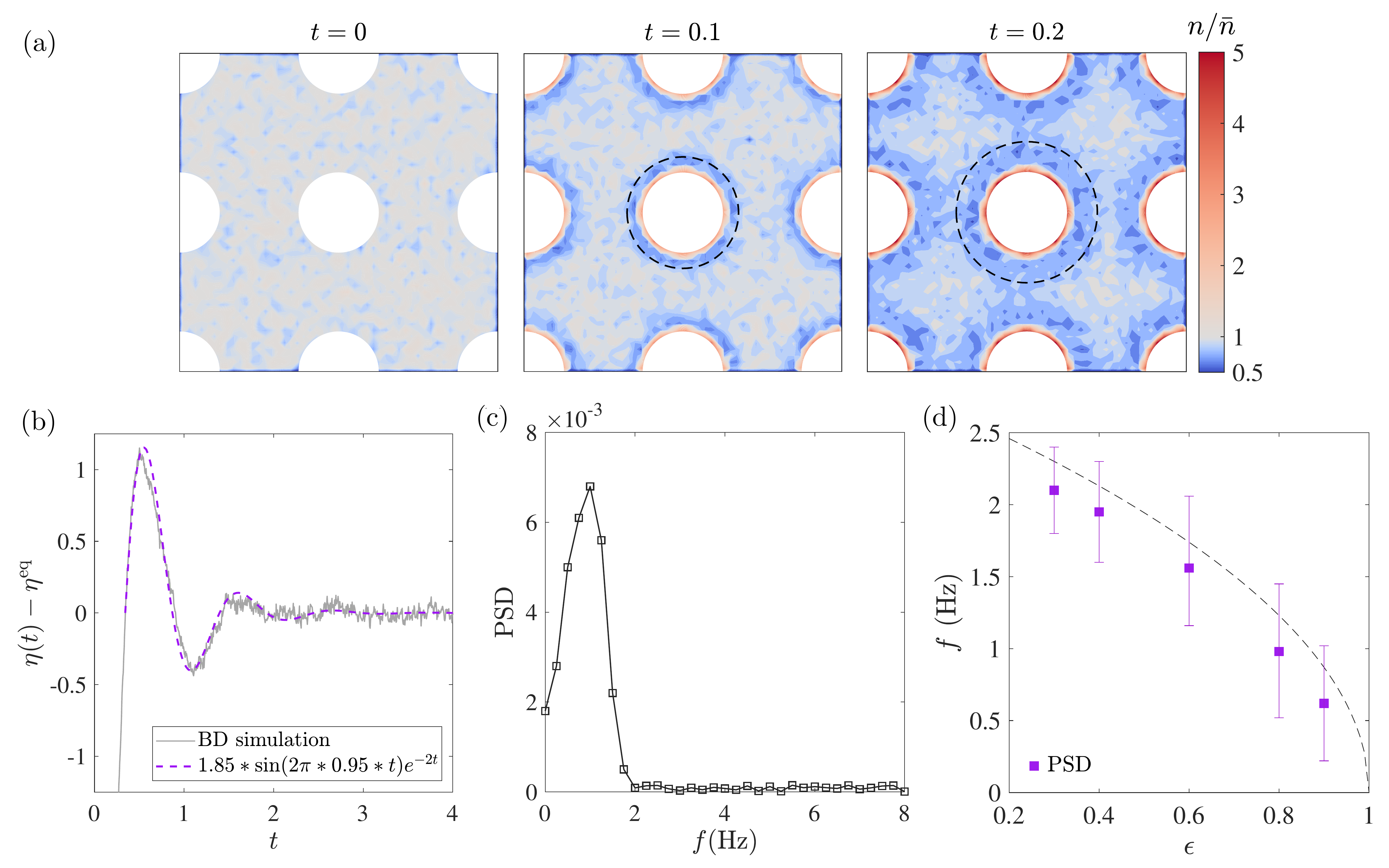}
\caption{(a) Snapshots from a BD simulation showing the evolution of the local number density, $n(\rb,t)$, of an initially homogeneous suspension ($n(\rb,0)=\bar{n}$) in a lattice of porosity $\epsilon=0.8$. Note the propagation of a depleted region ($n/\bar{n}<1$) away from pillars over time. The dashed line shows the instantaneous wavefront predicted using the theoretical wave velocity $c=\Pe_s/\sqrt{2}$. Parameters used in the simulation are: $\Pe_s=5, \kappa^2=0.1$, and $\bar{n}=2\times 10^6$. (b) Time evolution of $\eta(t)$ about its steady-state value $\eta^\mathrm{eq}$ shows damped oscillations, captured effectively using an exponentially decaying harmonic function. (c) Power Spectral Density (PSD) of the oscillatory part of $\eta(t)$ reveals the dominant frequencies. (d) The peak frequency determined from the PSD scales inversely with the size of the domain, $\sqrt{1-\epsilon}$, indicated by dashed line for reference.}
\label{fig:oscillations}
\end{figure}

The formation and propagation of a depleted region near a boundary is also observed in BD simulations in a periodic lattice (Fig. \ref{fig:oscillations}). At short times, particle accumulation on a pillar increases the surface number density, $\eta$. Higher $\Pe_s$ produces stronger accumulation, leaving behind a more pronounced depleted region. When the depletion wave from a neighboring pillar arrives, $\eta$ decreases temporarily, as more particles leave the pillar and enter the bulk; the magnitude of this drop is set by the ``strength" of the wave. Over time, some of these particles re-accumulate on the surface, producing a temporary increase in $\eta$ until the next depletion wave(s) arrives. Thus, the transient oscillations in the surface mean-field quantities noted in the main text can be understood as arising from the interaction of these depletion waves with boundaries. Since the depletion strength sets the amplitude of the oscillation, it increases with $\Pe_s$. Moreover, because the depletion strength decays as it propagates, stronger confinement (smaller $\epsilon$) can slightly increase the amplitudes. Another consequence of the long-time diffusive nature is that these oscillations are damped over time; an exponential decay fits the observation as shown in Fig. \ref{fig:oscillations}(b). The decay exponent is a found to predominantly depend on the activity, with higher $\Pe_s$ producing faster decay, consistent with our earlier observation that higher $\Pe_s$ equilibrates the system more rapidly (see Sec. \ref{sec:activity}).

In  a monodisperse square lattice, the closest surface-surface distance with neighboring pillars is $d=\ell/a-2$. The speed of the depletion front corresponds to that of any active matter wave in 2D, i.e. $c=\Pe_s/\sqrt{2}$ \cite{Dulaney20}, also validated by BD simulations (see Fig. \ref{fig:oscillations}(a)). 
Thus, the depletion fronts from the nearest pillars are expected to arrive at $t_1 = d/c$. When the characteristic size of the pillars is smaller than the distance separating them, i.e., $d \gtrsim 1$ (corresponding to $\epsilon \gtrsim 0.65)$, the mean surface number density drops as soon as the waves impact. Thus, in this case, $t_1$ would be an an estimate of the peak in the number density on the pillar. For the example shown in Fig. \ref{fig:oscillations}(b) (with $\epsilon=0.8\;, \Pe_s=5, \; \kappa^2=0.1$), $t_1=0.55$ indeed matches with the time of the observed maximum. 

The oscillations arise from a complex superposition of the depletion waves from the different pillars. But it has a dominant frequency which can be extracted by computing its power spectrum (Fig. \ref{fig:oscillations}(c)). This is because the waves decay rapidly, and only the first few surrounding pillars actually contribute to significant oscillations on the surface. Because the oscillations result from the ballistic nature of the particles, their frequencies scale linearly with the wave speed, $c \sim \Pe_s$ and inversely with the domain length, i.e. as $a/\ell \sim \sqrt{1-\epsilon}$ (see Fig. \ref{fig:oscillations}(d)).

\section{Brownian Dynamics simulations}
\label{app:B}

The instantaneous position of a particle $i$ in Cartesian coordinates is denoted by $\bv{R}_i=(x_i,y_i)$, and its orientation is $\pb_i=(\cos \theta_i, \sin \theta_i)$, where $\theta_i$ is the angle made by the director vector with respect to the $x$-axis. The change in the positions and orientations of the particles are determined using an Euler time-marching of Eqs.~\eqref{eq:Jeff1}-\eqref{eq:Jeff2}. Thus, during a time interval $\Delta t$, we have:
\begin{align}
        \Delta \bv{R}_i & = \Pe_s \pb_i\; \Delta t + \Pe_f\; \ub_i\; \Delta t + \sqrt{2\kappa^2 \Delta t}\; {\boldsymbol{\zeta}_T}, \\
        \Delta \theta_i & = \Pe_f (\pb_i \cdot \boldsymbol{\nabla} \ub_i \cdot \pb_i^{\perp}) \Delta t + \sqrt{2 \Delta t}\; {\zeta_R},
\end{align}
where $\pb_i^\perp=(-\sin \theta_i,\cos \theta_i)$, $\ub_i=\ub(\Rb_i)$ is the fluid velocity sampled at the position of the particle, and $\boldsymbol{\zeta}_T$ and $\zeta_R$ are independent Gaussian random variables with zero mean and unit variance. The velocity field and its gradient inside the unit cell are determined using the Method of Fundamental Solutions (see Appendix \ref{app:BIM}). For computing efficiency, both $\ub(\rb)$ and $\bnabla \ub(\rb)$ obtained after rescaling are pre-tabulated on a Cartesian grid in the rescaled cell, and linear interpolation is used to evaluate them at particle positions during simulations. The instantaneous numerical velocity of a particle in a given time step is then $\vb_i = \Delta \Rb_i/\Delta t$. If the particle spuriously enters the pillar, the component of its velocity normal to the local surface of the pillar is set to zero for that step $(\vb_i \cdot \nb = 0)$, and the displacement is recalculated. This approach is effective in the case of smooth convex geometries. We chose an integration step $\Delta t$ depending on the porosity of the medium; typically $\Delta t \sim O(10^{-4})-O(10^{-3})$. Finally, we validated the results of our simulations by measuring the dispersivity of the particles in a simple shear flow, for which analytical solutions exist \cite{tenHagen11,Sandoval16}. Additionally, we validated the dispersion through a porous matrix by comparing against published results for the same system \cite{AlonsoMatilla19}.

\section{Method of Fundamental Solutions to determine the flow field}
\label{app:BIM}

\begin{figure}[t]
\centering
\includegraphics[width=\textwidth]{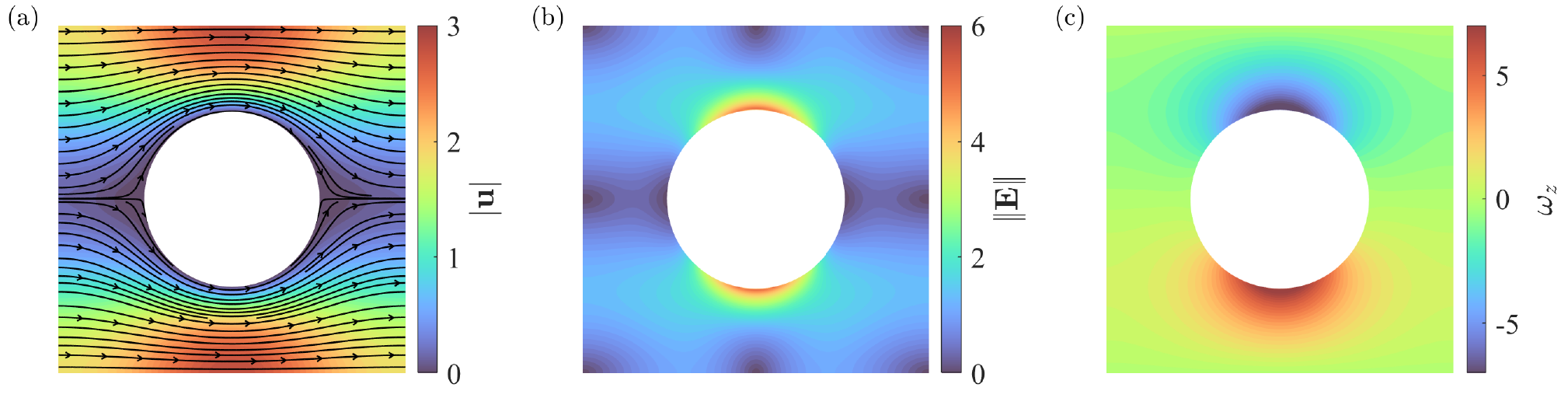} 
\caption{Flow field around a pillar of unit radius in a doubly-periodic cell with porosity $\epsilon=0.8$ due to a pressure drop imposed between the left and right boundaries. Shown are (a) the dimensionless flow velocity $\ub$, (b) the norm of dimensionless strain-rate tensor, $\boldsymbol{\mathsf{E}} = (\bnabla\ub + (\bnabla\ub)^\top)/2$, and (c) the dimensionless out-of-plane vorticity, $\omega_z=\hat{\boldsymbol{z}} \cdot (\bnabla\times\ub)$.}
\label{fig:flow_field}
\end{figure}

Fluid flow is introduced in the lattice by an externally-imposed pressure jump across the left and right sides of the doubly-periodic unit cell, i.e., along the $x$-axis. The Stokes flow field is calculated numerically using the {Method of Fundamental Solutions} (MFS) routine developed in \cite{Barnett15, Barnett18}. The spectrally accurate scheme overcomes the need to use periodic Green's functions for the computation of the hydrodynamic fields. A close-evaluation quadrature is used to maintain the numerical accuracy even in cases of low porosity where the boundaries of the pillars are very close to each other. The flow calculation is performed in dimensionless variables inside a square periodic cell of unit edge length. An example of the flow field when a pressure jump $[\![p]\!]_x$ is applied between the left and right edges of the doubly-periodic cell containing a single pillar is shown in Fig.~\ref{fig:flow_field}.

We used a two-pronged approach to validate the numerical solution. The first was by noting that in the absence of inertia, the total drag force on the pillar ($F_d$) must balance the force from the applied pressure jump across the unit cell ($F_p = [\![p]\!]_x$). We found that the difference between these two quantities is only of the order of the machine error $|F_d-F_p| \sim \mathcal{O}(10^{-17})$ when performing boundary integrals with an appropriate number of collocation points on the surface. The second validation was done by comparing the dimensionless drag with that computed using a highly-accurate lattice sums approach by Greengard \& Kropinski \cite{Greengard04}. In our particular case, the net drag is along the $x$-direction \cite{Sangani82} and is given in dimensionless form by
\begin{equation}
    \mathcal{D}_x = \frac{[\![p]\!]_x}{J_x} \quad \mathrm{with} \quad J_x = \int_{B_L} \ub(\xb) \cdot \hat{\xb}\; \mathrm{d}y,
    \label{eq:drag}
\end{equation}
where $\ub(\xb)$ is the velocity field determined using MFS, and the flux $J_x$ is evaluated at the left (or right) unit cell boundary. In the case of a single pillar, symmetry about the $x$-axis implies that there is no net flux in the $y$-direction, i.e. $J_y=0$ here. In practice, the flux $\Jb = (J_x,J_y)$ is determined using a more direct relation it has with the fundamental solutions at the pillar surface \cite{Barnett18}. The velocity field calculated using MFS was then scaled by $|\Jb|$ for use in the BD simulations (in Fig.~\ref{fig:flow_field}).  Note that in the Stokes regime the structure of the flow field and the dimensionless drag on a pillar are geometric properties that depend neither on the physical properties of the fluid nor the applied external forcing. The computations performed here were able to reproduce all the decimal places of $\mathcal{D}_x$ determined by \cite{Greengard04} for all porosities.

\section{Boundary layer on unbounded pillar in the limit of large activity}
\label{app:approx}

When $\Pe_s \gg 1$, along with the assumption that $\kappa \ll 1$, the boundary layer thickness in Eq.~\eqref{eq:delta} approximates to 
\begin{equation}
    \delta \approx \frac{\sqrt{2}\; \kappa^2}{\Pe_s} \ll 1.
\end{equation}
More precisely, the above approximation holds true when $\Pe_s/\kappa \gg \sqrt{2}$. In this asymptotic limit, the modified Bessel functions in Eq.~\eqref{eq:YanBradyA} for the particle density outside of an isolated pillar can expanded as Taylor series to yield the approximation 
\begin{align}
    \frac{\eq{n}(r)}{n_\infty} \approx 1 + \frac{e^{(1-r)/\delta} r^{-1/2}\left(1-\frac{\delta}{8 r}+ O(\delta^2)\right)}{\delta\left(1+\frac{3\delta}{8} + O(\delta^2)\right) + \frac{\delta^2}{\kappa^2}\left(1+\frac{7\delta}{8} + O(\delta^2) \right)} .
    \label{eq:approxn}
\end{align}
Thus, for large activity, the number density of particles at the pillar surface, $\eq{n_s}=\eq n(r=1)$, approximates to
\begin{align}
    \frac{\eq{n}_s}{n_\infty}  \approx 1 + \frac{1}{\kappa^2}\left(\frac{\Pe_s}{\sqrt{2}}-1\right).
\end{align}
The exact expression for the boundary layer-averaged number density in Eq.~\eqref{eq:BLmean} is
\begin{equation}
    \frac{\eq{\nA}}{n_\infty} = 1+ \frac{4 (\Pe_s/\kappa)^2 (K_1(1/\delta)-(1+\delta)K_1(1+1/\delta))}{(\delta+2)(K_0(1/\delta)\;(2-(\Pe_s/\kappa)^2) + K_2(1/\delta)\;(2+(\Pe_s/\kappa)^2))}.
    \label{eq:blavg}
\end{equation}
Expanding this expression in a Taylor series yields the approximation
\begin{equation}
    \frac{\eq{\nA}}{n_\infty} \approx 1+\frac{0.632}{\kappa^2}\left(\frac{\Pe_s}{\sqrt{2}}-1\right).
\end{equation}

The steady-state polarization field near an unbounded pillar is obtained using the Fokker-Planck equation
\begin{equation}
    \bnabla \cdot \bv{j}_n=0 \implies \bnabla \cdot \mb - \frac{\delta}{\sqrt{2}} \nabla^2 n = 0,
\end{equation}
with the normal density flux vanishing on the pillar surface: $\nb \cdot \bv{j}_n (r=1) = 0$. For a radially-symmetric system, this implies that
\begin{equation}
    \frac{\eq{m}}{n_\infty} = \frac{\delta}{\sqrt{2}} \pard{}{r}\left(\frac{n(r)}{n_\infty}\right),
    \label{eq:approxm}
\end{equation}
where $\eq m$ is the steady-state radial polarization. Substituting \eqref{eq:approxn} into \eqref{eq:approxm}, we obtain asymptotic approximations for the surface and the boundary-averaged values of the radial polarization, 
\begin{equation}
    \frac{\eq{m}_s}{\eq n_s} \approx \frac{-1}{\sqrt{2}}+\frac{\kappa^2}{2\;\Pe_s},\quad \mathrm{and} \quad \eq{\mu}=\frac{\eq\mA}{\eq\nA} \approx \frac{-1}{\sqrt{2}}+\frac{1.582\; \kappa^2}{\Pe_s}.
    \label{eq:limit_mu1}
\end{equation}
respectively. Note that as $\Pe_s \to \infty$, the surface radial polarization converges to a constant value of $\eq\mu \to -1/\sqrt{2}$.



\bibliography{biblio2.bib}

\end{document}